		\documentclass[10pt,journal,compsoc]{IEEEtran}
		\ifCLASSOPTIONcompsoc
		  \usepackage[no compress]{cite}
		\else
		  \usepackage{cite}
		\fi
		\ifCLASSINFOpdf
		\else
		\fi
		\usepackage{mathtools}
		\usepackage{stmaryrd}
		\usepackage{amssymb}
		\usepackage{amsmath}
		\usepackage{moreverb}
		\usepackage{amssymb}
		\usepackage{lineno}
		\usepackage{graphicx}
		\usepackage{amssymb}
		\usepackage{amsmath}
		\usepackage{moreverb}
		\usepackage{amssymb}
		\usepackage{lineno}
		\usepackage{graphicx}
		\graphicspath{ {images/} }
		\usepackage{caption}
		\usepackage{subcaption}
		\usepackage{algpseudocode}
		\usepackage{multirow}
		\usepackage[utf8]{inputenc}
		\usepackage[english]{babel}
		\usepackage{mathtools}
		\usepackage[colorlinks,bookmarksopen,bookmarksnumbered,citecolor=red,urlcolor=red]{hyperref}
		\usepackage{rotating,booktabs,comment,array,tabu,pdflscape}
		\usepackage{longtable,slashbox,color,colortbl}
		\usepackage{float}
		\usepackage{stmaryrd}
		\usepackage{multirow}
		\usepackage{epstopdf}
		\usepackage{makecell}
		\usepackage{multirow}
		\usepackage{booktabs}
		\usepackage{graphicx}
		\usepackage[table,xcdraw]{xcolor}
		\definecolor{Gray}{gray}{0.9}
		\usepackage[linesnumbered,boxed,lined, ruled]{algorithm2e}
		\let\oldnl\nl% Store \nl in \oldnl
		\newcommand{\nonl}{\renewcommand{\nl}{\let\nl\oldnl}}% Remove line number for one line
		\usepackage{enumitem}
		\usepackage{tabularx}
		\usepackage{ltablex} 
		\usepackage{longtable}
		\usepackage{supertabular}
		\usepackage{color, colortbl}
		\usepackage{csquotes}
		\usepackage{subfloat}
		\usepackage[bottom]{footmisc}
		
		\usepackage{stackengine}
		
		\usepackage{amsmath}

\begin{document}
			\fontsize{9.5pt}{11.5pt} 
			\title{A Distributed Deep Reinforcement Learning Technique for Application Placement in Edge and Fog Computing Environments}

		\author{Mohammad~Goudarzi,~\IEEEmembership{Member,~IEEE,}
			Marimuthu~Palaniswami,~\IEEEmembership{Fellow,~IEEE,}
			and~Rajkumar~Buyya,~\IEEEmembership{Fellow,~IEEE}% <-this % stops a space
			\IEEEcompsocitemizethanks{\IEEEcompsocthanksitem 
				M. Goudarzi and R. Buyya are with the Cloud Computing and Distributed Systems (CLOUDS) Laboratory, School of Computing and Information Systems, The University of Melbourne, Australia.
				
				\IEEEcompsocthanksitem
				M. Palaniswami is with the Department of Electrical and Electronic Engineering, The University of Melbourne, Australia \protect\\
				E-mail: mgoudarzi@student.unimelb.edu.au, palani@unimelb.edu.au, rbuyya@unimelb.edu.au. 
				
				\IEEEcompsocthanksitem Corresponding Author: M. Goudarzi 
				 
				%\IEEEcompsocthanksitem 
			}% <-this % stops an unwanted space
		}
			

			\markboth{IEEE TRANSACTIONS ON MOBILE COMPUTING, VOL.XX, NO.XX, 202X}%
			{Shell \MakeLowercase{\textit{et al.}}: Bare Demo of IEEEtran.cls for Computer Society Journals}

			\IEEEtitleabstractindextext{
				%\IEEEdisplaynontitleabstractindextext
				\begin{abstract}
					Fog/Edge computing is a novel computing paradigm supporting resource-constrained Internet of Things (IoT) devices by placement of their tasks on edge and/or cloud servers. Recently, several Deep Reinforcement Learning (DRL)-based placement techniques have been proposed in fog/edge computing environments, which are only suitable for centralized setups. The training of well-performed DRL agents requires manifold training data while obtaining training data is costly. Hence, these centralized DRL-based techniques lack generalizability and quick adaptability, thus failing to efficiently tackle application placement problems. Moreover, many IoT applications are modeled as Directed Acyclic Graphs (DAGs) with diverse topologies. Satisfying dependencies of DAG-based IoT applications incur additional constraints and increase the complexity of placement problem. To overcome these challenges, we propose an actor-critic-based distributed application placement technique, working based on the IMPortance weighted Actor-Learner Architectures (IMPALA). IMPALA is known for efficient distributed experience trajectory generation that significantly reduces exploration costs of agents. Besides, it uses an adaptive off-policy correction method for faster convergence to optimal solutions. Our technique uses recurrent layers to capture temporal behaviors of input data and a replay buffer to improve the sample efficiency. The performance results, obtained from simulation and testbed experiments, demonstrate that our technique significantly improves execution cost of IoT applications up to 30\% compared to its counterparts.
					
        		\end{abstract}
				
				}

			\maketitle

			\IEEEdisplaynontitleabstractindextext

			% \fi
			\IEEEpeerreviewmaketitle

			\IEEEraisesectionheading{\section{Introduction}\label{sec:introduction}} 
			
			\IEEEPARstart{I}{n} recent years, new computing and communication technologies have rapidly advanced, leading to the proliferation of smart Internet of Things (IoT) devices (e.g., sensors, smartphones, cameras, vehicles) \cite{hu2017survey}. These advancements empower IoT devices to run a multitude of resource-hungry and latency-sensitive IoT applications. These emerging IoT applications increasingly demand computing, storage, and communication resources for the execution \cite{wang2019Taskdelay}. Also, the execution of such resource-hungry applications requires a significant amount of energy consumption. Hence, limited computing, storage, and battery capacity of IoT devices, directly affect the performance of IoT applications and user experience \cite{goudarzi2020application}.   
			\par
			The Cloud computing paradigm, as a centralized solution, is one of the main enablers of the IoT, providing unlimited and elastic remote computing and storage resources for the execution of computation-intensive IoT applications \cite{deng2020optimal}. All/some computation-intensive constituent parts (e.g., service, modules, tasks) of IoT applications can be placed (i.e., offloaded) on remote Cloud Servers (CSs) for execution and storage in order to reduce the execution time of IoT applications and energy consumption of IoT devices \cite{wang2019delay,goudarzi2017fast}. However, due to low bandwidth and high communication latency between IoT devices and CSs, the requirements of latency-sensitive IoT applications cannot be efficiently satisfied \cite{xu2019computation}. Besides, low bandwidth and high latency of CSs may incur more energy consumption for IoT devices due to higher active communication time with CSs. To improve the high communication latency and low bandwidth of CSs, fog computing paradigm, as a distributed solution, has emerged. In fog computing, heterogeneous Fog Servers (FSs) are distributed in the proximity of IoT devices, through which IoT devices can access the computing and storage resources with higher bandwidth and less communication latency, compared to CSs \cite{goudarzi2019fog}. However, these FSs usually have limited resources (e.g., CPU, RAM) in comparison to CSs. In our view, edge computing harnesses only distributed edge resources at the proximity of IoT devices while fog computing harnesses both edge and cloud resources to address the requirements of both computation-intensive and latency-sensitive IoT applications (although some works use these terms interchangeably). 
			\par
			In real-world scenarios, many IoT applications (e.g., face recognition \cite{ra2011odessa}, smart healthcare \cite{gia2015fog}, and augmented reality \cite{al2017energy}) are modeled as a Directed Acyclic Graph (DAG), in which nodes and edges represent tasks and data communication among dependent tasks, respectively. These DAG-based IoT applications incur higher complexity and constraints when making placement decisions for the execution of IoT applications. Hence, placement/offloading of IoT applications, comprised of dependent tasks, on/to suitable servers with the minimum execution time and energy consumption is an important and yet challenging problem in fog computing. Many heuristics, approximation, and rule-based solutions are proposed for this NP-hard problem \cite{brogi2017qos, bittencourt2017mobility,goudarzi2016mobile}. Although these techniques work well in general cases, they heavily rely on comprehensive knowledge about the IoT applications and resource providers (e.g., CSs or FSs). The fog computing environment is stochastic in several aspects, such as arrival rate of application placement requests, dependency among tasks, number of tasks per IoT application, resource requirements of applications, and available remote resources, just to mention a few. Therefore, heuristic-based techniques cannot efficiently adapt to constant changes in the fog computing environments \cite{jeff2018ml}.
			\par
			 Deep Reinforcement Learning (DRL) provides a promising solution by combining Reinforcement Learning (RL) with Deep Neural Network (DNN). Since DRL agents can accurately learn the optimal policy and long-term rewards without prior knowledge of the system \cite{wang2020fast}, they help solve complex problems in dynamic and stochastic environments such as fog computing, especially when the state space is so large \cite{jeff2018ml,mao2018deep}.
			 Although the effectiveness of DRL techniques is shown in several works \cite{huang2019deep,chen2018optimized,huang2019deepDigital,lu2020edge,sun2019application}, there are yet several challenges for practical realizations of these techniques in fog computing environments. In DRL, the agent interacts with the environment using trial and error (i.e., exploration) and records the trajectories of experiences (i.e., sequences of states, actions, and rewards) in large quantities with high diversity. These experience trajectories are used to learn the optimal policy in the training phase. In complex environments, such as fog computing, DRL agents require a large number of interactions with the environment to obtain sufficient trajectories of experience to capture the properties of the environment. Therefore, the exploration cost of agents increases. Obviously, it negatively affects the user experience in the fog computing environment, because the training of the DRL agents in such complex environments is a time-consuming process. The centralized DRL agents used in fog computing environments are not suitable for the highly distributed and stochastic environments \cite{tuli2020dynamic}. Hence, a key problem is how to adapt distributed DRL techniques to efficiently perform in fog computing environments. Considering the distributed nature of fog computing environments, the application placement engines can be placed on different FSs, that work in parallel and efficiently produce diverse experience trajectories with less exploration costs. However, other challenges may arise such as how these trajectories can be efficiently and practically used to learn the optimal policy.       
			\par
			To address the aforementioned challenges, we propose an EXperience-sharing Distributed Deep Reinforcement Learning-based application placement technique, called $X$-DDRL, to efficiently capture complex dynamics of DAG-based IoT applications and FSs' resources. The $X$-DDRL uses IMPortance weighted Actor-Learner Architectures (IMPALA), proposed by Espeholt et al. ~\cite{ espeholt2018impala}, which is a distributed DRL agent that uses an actor-learner framework to learn the optimal policy. In IMPALA, several actors interact with the environments in parallel and produce diverse experience trajectories in a timely manner. Then, these experience trajectories are periodically forwarded to the learner for the training and learning of the optimal policy. After each policy update of the learner, actors reset their parameters with the learner's one and independently continue their explorations. As a result of this distributed and collaborative experience-sharing between actors and learners, the exploration costs reduce significantly, and the experience trajectories are efficiently reused. However, due to decoupled acting and learning, a policy gap between actors and learners arises, which can be corrected by V-trace off-policy correction method \cite{espeholt2018impala}. Moreover, we use Recurrent Neural Networks (RNN) to accurately identify the temporal patterns across different features of the input. Finally, the $X$-DDRL uses experience replay to break the strong correlation between generated experience trajectories and improve sample efficiency. 
			\par
			The main contributions of this paper are summarized as follows:
			 %\vspace{-1mm}
			\begin{itemize}
				\item A weighted cost model for application placement of DAG-based IoT applications is proposed to minimize the execution time of IoT applications and energy consumption of IoT devices. Then, this weighted cost model is adapted to be used in DRL-based techniques.
				\item A pre-scheduling technique is put forward to define an execution order for dependent tasks within each DAG-based IoT application.
				\item We propose a dynamic and distributed DRL-based application placement technique for complex and stochastic fog computing environments, working based on the IMPALA framework. Our technique uses RNN to capture complex patterns across different features of the input. Moreover, it uses an experience replay buffer which remarkably helps sampling efficiency and breaks the strong correlation between experience trajectories.
				\item We conduct simulation and testbed experiments using a wide range of synthetic DAGs, derived from the real-world IoT applications, to cover diverse application dependency models, task numbers, and execution costs. Also, the performance of our technique is compared with two state-of-the-art DRL techniques, called Double Deep Q Learning (Double-DQN), and Proximal Policy Optimization (PPO), and a greedy-based heuristic.
			\end{itemize}  
			\par
			The rest of the paper is organized as follows. Relevant DRL-based application placement techniques in edge and fog computing environments are discussed in Section~\ref{relatedw}. The system model and problem formulations are presented in Section~\ref{system}. Section~\ref{sec:DRLModel} describes the DRL-based model and its main concepts. Section~\ref{placement} presents our proposed distributed DRL-based application placement framework. We evaluate the performance of our technique and compare it with state-of-the-art techniques in Section~\ref{evaluation}. Finally, Section~\ref{conclusion} concludes the paper and draws future works.
			
			%\IEEEpubidadjcol
			%\vspace{-0.2cm}
			\section{Related Work}
			\label{relatedw}
			Considering the large number of works in application placement techniques, in this section, related works for DRL-based application placement techniques in fog/edge computing environments are studied. However, detailed related works for the non-learning-based application placement techniques and frameworks are available in \cite{goudarzi2020application,goudarzi2021distributed,deng2021fogbus2}.   
			\par
			 DRL-based works are first divided into edge computing and fog computing. Edge computing works only consider the resources in the proximity of IoT users while fog computing ones take advantage of both edge resources and remote cloud resources. Hence, the heterogeneity of resources is higher in the fog computing works, which leads to higher complexity for DRL-based application placement techniques to identify the features of the environments. Besides, works are further categorized into independent and dependent categories based on the dependency model of their IoT applications' granularity (e.g., tasks, modules). In IoT applications with dependent tasks (i.e., DAGs), each task can be executed only when its parent tasks finish their execution, while tasks of independent IoT applications do not have such constraints for execution. Therefore, works in the dependent category have more constraints, and hence the DRL agent requires specific considerations compared to works in the independent category to efficiently learn the optimal policy.    
			%\vspace{-3mm}
			\subsection{Edge Computing}
			In the independent category, Huang et al.~\cite{huang2018distributed} proposed a DRL-based offloading algorithm to minimize the system cost, in which parallel computing is used to speed up the computation of a single edge server. Min et al.~\cite{min2019learning} proposed a fast deep Q-network (DQN) based offloading scheme, combining the deep learning and hotbooting techniques to improve the learning speed of Q-learning. Huang et al.~\cite{huang2019deep} proposed a quantized-based DRL method to optimize the system energy consumption for faster processing of IoT devices' requests. Chen et al.~\cite{chen2018optimized} proposed a double DQN-based algorithm to minimize the energy consumption and execution time of independent tasks of IoT applications. Huang et al.~\cite{huang2019deepDigital} also proposed a DRL-based offloading framework based on DQN that jointly considers offloading decisions and resource allocations. Chen et al.~\cite{chen2019iraf} proposed a joint offloading framework with DRL to make an offloading decision based on the information of applications' tasks and network conditions where the training data is generated from the searching process of the Monte Carlo tree search algorithm. Lu et al.~\cite{lu2020edge} proposed a Deep Deterministic Policy Gradients (DDPG)-based algorithm for computation offloading of multiple IoT users to a single edge server to improve the quality of experience of users. To improve the convergence of the DQN algorithm in an edge computing environment, Xiong et al.~\cite{xiong2020resource} proposed a DQN-based algorithm combined with multiple replay memories to minimize the execution time of one IoT application. Qiu et al.~\cite{qiu2020distributed} studied the distributed DRL in an edge computing environment with a single edge server to minimize the energy cost of running IoT applications, consisted of independent tasks. To obtain this goal, they combined deep neuro-evolution and policy gradient to improve the convergence results.
			\par
			In the dependent category, Wang et al.~\cite{wang2020fast} proposed a meta reinforcement learning algorithm based on the Proximal Policy Optimization (PPO). The main goal of this work is to minimize the execution time of dependent IoT applications, situated in the proximity of a single edge server.
			
			\subsection{Fog Computing}
			In the independent category, Gazori et al.~\cite{gazori2020saving} targeted task scheduling of independent IoT applications to minimize long-term service delay and system cost. To obtain this, they used a double DQN-based scheduling algorithm combined with an experience replay buffer. Tuli et al.~\cite{tuli2020dynamic} proposed Asynchronous-Advantage-Actor-Critic (A3C) learning-based technique combined with Recurrent Neural Network (RNN) for the scheduling of independent IoT applications to minimize total system cost.
			\par
			In the dependent category, Lu et al.~\cite{lu2020optimization} proposed a DQN-based algorithm to minimize the overall system cost. Although they consider dependencies among constituent parts of each IoT application, they only consider the sequential dependency model among tasks of an IoT application, where there are no tasks for parallel execution.
			\subsection{A Qualitative Comparison}

			\begin{table*}[!ht]
				\footnotesize
				\centering
				\caption{A qualitative comparison of related works with ours}
				\label{tab:relatedwork}
				
				\resizebox{1\textwidth}{!}{%
				\renewcommand{\arraystretch}{1.5}
				\footnotesize
\begin{tabular}{|c|c|c|c|c|c|c|c|c|c|c|c|c|c|c|} 
	\hline
	\multirow{5}{*}{Techniques}                                                                                & \multirow{5}{*}{Category}                                                  & \multicolumn{3}{c|}{\multirow{2}{*}{Application Properties}}                                                                            & \multicolumn{5}{c|}{\multirow{2}{*}{Architectural Properties}}                                                                                             & \multicolumn{5}{c|}{Application Placement Engine Properties}                                                                                                                                                                             \\ 
	\cline{11-15}
	&                                                                            & \multicolumn{3}{c|}{}                                                                                                                   & \multicolumn{5}{c|}{}                                                                                                                                      & \multirow{4}{*}{\begin{tabular}[c]{@{}c@{}} Main\\Method \end{tabular}}                                  & \multirow{4}{*}{\begin{tabular}[c]{@{}c@{}} Task\\Priority \end{tabular}} & \multicolumn{3}{c|}{Decision Parameters}                                              \\ 
	\cline{3-10}\cline{13-15}
	&                                                                            & \multirow{3}{*}{Dependency} & \multirow{3}{*}{\begin{tabular}[c]{@{}c@{}}Task\\Number\end{tabular}} & \multirow{3}{*}{Heterogeneity} & \multicolumn{2}{c|}{IoT Device Layer}                                         & \multicolumn{2}{c|}{Edge/Fog Layer}         & \multirow{3}{*}{\begin{tabular}[c]{@{}c@{}} Multi\\Cloud \end{tabular}}  &                                                               &                                                                           & \multirow{3}{*}{Time}      & \multirow{3}{*}{Energy}    & \multirow{3}{*}{Weighted}   \\ 
	\cline{6-9}
	&                                                                            &                              &                                                                         &                                & Number               & \begin{tabular}[c]{@{}c@{}}Request\\Type \end{tabular} & Number               & Heterogeneity        &                              &                                                               &                                                                           &                            &                            &                             \\ 
	\hline
	\begin{tabular}[c]{@{}c@{}}\cite{huang2018distributed}\end{tabular}          & \multirow{12}{*}{\begin{tabular}[c]{@{}c@{}}Edge\\Computing \end{tabular}} & \multirow{11}{*}{Independent} & Multiple                                                                & Heterogeneous                  & Multiple             & Homogeneous                                            & Single               & Homogeneous          & $\times$    & \_                                                            & $\times$                                                 & $\times$  & $\times$  & \checkmark   \\ 
	\cline{1-1}\cline{4-15}
	\begin{tabular}[c]{@{}c@{}} \cite{min2019learning}\end{tabular}                &                                                                            &                              & Single                                                                  & Homogeneous                    & Multiple             & Heterogeneous                                          & Single               & Homogeneous          & $\times$    & \_                                                            & $\times$                                                 & $\times$  & \checkmark  & $\times$   \\ 
	\cline{1-1}\cline{4-15}
	\begin{tabular}[c]{@{}c@{}}\cite{huang2019deep}\end{tabular}                  &                                                                            &                              & Single                                                                  & Homogeneous                    & Single               & Homogeneous                                            & Multiple             & Heterogeneous        & $\times$    & DQN                                                           & $\times$                                                 & $\times$  & \checkmark  & $\times$   \\ 
	\cline{1-1}\cline{4-15}
	\begin{tabular}[c]{@{}c@{}}\cite{chen2018optimized}\end{tabular}     &                                                                            &                              & Single                                                                  & Heterogeneous                  & Multiple             & Heterogeneous                                          & Single               & Homogeneous          & $\times$    & \begin{tabular}[c]{@{}c@{}}Double\\DQN\end{tabular}           & $\times$                                                 & \checkmark & \checkmark & $\times$   \\ 
	\cline{1-1}\cline{4-15}
	\begin{tabular}[c]{@{}c@{}}\cite{huang2019deepDigital}\end{tabular} &                                                                            &                              & Multiple                                                                & Heterogeneous                  & Multiple             & Heterogeneous                                          & Single               & Homogeneous          & $\times$    & DQN                                                           & $\times$                                                 &\checkmark & \checkmark & \checkmark  \\ 
	\cline{1-1}\cline{4-15}
	\begin{tabular}[c]{@{}c@{}} \cite{chen2019iraf}\end{tabular}                    &                                                                            &                              & Single                                                                  & Heterogeneous                  & Multiple             & Heterogeneous                                          & Multiple             & Heterogeneous        & $\times$    & MCTS                                                          & $\times$                                                 & \checkmark  & \checkmark  & $\times$   \\ 
	\cline{1-1}\cline{4-15}
	\begin{tabular}[c]{@{}c@{}} \cite{lu2020edge}\end{tabular}                     &                                                                            &                              & Multiple                                                                & Heterogeneous                  & Multiple             & Heterogeneous                                          & Single               & Homogeneous          & $\times$    & DDPG                                                          & $\times$                                                 & \checkmark  & \checkmark  & \checkmark   \\ 
	\cline{1-1}\cline{4-15}
	\begin{tabular}[c]{@{}c@{}}\cite{xiong2020resource}\end{tabular}             &                                                                            &                              & Multiple                                                                & Heterogeneous                  & Multiple             & Homogeneous                                            & Single               & Homogeneous          & $\times$    & DQN                                                           & $\times$                                                 & \checkmark  & $\times$  & $\times$   \\ 
	\cline{1-1}\cline{4-15}
	\begin{tabular}[c]{@{}c@{}}\cite{qiu2020distributed}\end{tabular}               &                                                                            &                              & Multiple                                                                & Heterogeneous                  & Multiple             & Heterogeneous                                          & Single            & Homogeneous        & $\times$    & \begin{tabular}[c]{@{}c@{}}Deep\\Neuroevolution \end{tabular} & $\times$                                                 & $\times$  &  \checkmark  & $\times$   \\ 
	\cline{1-1}\cline{3-15}
	\begin{tabular}[c]{@{}c@{}}\cite{wang2020fast}\end{tabular}                       &                                                                            & Dependent                    & Multiple                                                                & Heterogeneous                  & Multiple             & Heterogeneous                                          & Single               & Homogeneous          & $\times$    & PPO                                                           & \checkmark                                                 & \checkmark  & $\times$  & $\times$   \\ 
	\hline

%	\cline{1-1}\cline{4-15}
			\begin{tabular}[c]{@{}c@{}}\cite{gazori2020saving}\end{tabular}           &  \multirow{4}{*}{\begin{tabular}[c]{@{}c@{}}Fog \\Computing \end{tabular}}                                                                          & \multirow{2}{*}{Independent}                              & Multiple                                                                & Heterogeneous                  & Multiple             & Heterogeneous                                          & Multiple             & Heterogeneous        & $\times$    & \begin{tabular}[c]{@{}c@{}}Double\\DQN \end{tabular}          & $\times$                                                 & \checkmark & $\times$  & $\times$   \\ 
			\cline{1-1}\cline{4-15}
			\begin{tabular}[c]{@{}c@{}}\cite{tuli2020dynamic}\end{tabular}                      &                                                                            &                              & Multiple                                                                & Heterogeneous                  & Multiple             & Heterogeneous                                          & Multiple             & Heterogeneous        & $\times$    & A3C                                                           & $\times$                                                 & \checkmark  & \checkmark  & \checkmark   \\ 
			\cline{1-1}\cline{3-15}
			\begin{tabular}[c]{@{}c@{}}\cite{lu2020optimization}\end{tabular}           &                                                                            & \multirow{3}{*}{Dependent}   & Multiple                                                                & Heterogeneous                  & Multiple             & Homogeneous                                            & Multiple             & Heterogeneous        & $\times$    & DQN                                                           & $\times$                                                 & $\times$  & \checkmark  & $\times$   \\ 
			\cline{1-1}\cline{4-15}
			\begin{tabular}[c]{@{}c@{}}$X$-DDRL \end{tabular}                                                &                                                                            &                              & Multiple                                                                & Heterogeneous                  & Multiple             & Heterogeneous                                          & Multiple             & Heterogeneous        & \checkmark    & IMPALA                                                        & \checkmark                                                 & \checkmark  & \checkmark  & \checkmark   \\ 
			\hline
			\multicolumn{1}{l}{}                                                                                       & \multicolumn{1}{l}{}                                                       & \multicolumn{1}{l}{}         & \multicolumn{1}{l}{}                                                    & \multicolumn{1}{l}{}           & \multicolumn{1}{l}{} & \multicolumn{1}{l}{}                                   & \multicolumn{1}{l}{} & \multicolumn{1}{l}{} & \multicolumn{1}{l}{}         & \multicolumn{1}{c}{}                                          & \multicolumn{1}{c}{}                                                      & \multicolumn{1}{l}{}       & \multicolumn{1}{l}{}       & \multicolumn{1}{l}{}        \\
			\multicolumn{1}{l}{}                                                                                       & \multicolumn{1}{l}{}                                                       & \multicolumn{1}{l}{}         & \multicolumn{1}{l}{}                                                    & \multicolumn{1}{l}{}           & \multicolumn{1}{l}{} & \multicolumn{1}{l}{}                                   & \multicolumn{1}{l}{} & \multicolumn{1}{l}{} & \multicolumn{1}{l}{}         & \multicolumn{1}{c}{}                                          & \multicolumn{1}{c}{}                                                      & \multicolumn{1}{l}{}       & \multicolumn{1}{l}{}       & \multicolumn{1}{l}{}        \\
			\multicolumn{1}{c}{}                                                                                       & \multicolumn{1}{c}{}                                                       & \multicolumn{1}{c}{}         & \multicolumn{1}{c}{}                                                    & \multicolumn{1}{c}{}           & \multicolumn{1}{c}{} & \multicolumn{1}{l}{}                                   & \multicolumn{1}{c}{} & \multicolumn{1}{l}{} & \multicolumn{1}{c}{}         & \multicolumn{1}{c}{}                                          & \multicolumn{1}{c}{}                                                      & \multicolumn{1}{c}{}       & \multicolumn{1}{c}{}       & \multicolumn{1}{c}{}        \\
			\multicolumn{1}{c}{}                                                                                       & \multicolumn{1}{c}{}                                                       & \multicolumn{1}{c}{}         & \multicolumn{1}{c}{}                                                    & \multicolumn{1}{c}{}           & \multicolumn{1}{c}{} & \multicolumn{1}{l}{}                                   & \multicolumn{1}{c}{} & \multicolumn{1}{l}{} & \multicolumn{1}{c}{}         & \multicolumn{1}{c}{}                                          & \multicolumn{1}{c}{}                                                      & \multicolumn{1}{c}{}       & \multicolumn{1}{c}{}       & \multicolumn{1}{c}{}        \\
			\multicolumn{1}{l}{}                                                                                       & \multicolumn{1}{l}{}                                                       & \multicolumn{1}{l}{}         & \multicolumn{1}{l}{}                                                    & \multicolumn{1}{l}{}           & \multicolumn{1}{l}{} & \multicolumn{1}{l}{}                                   & \multicolumn{1}{l}{} & \multicolumn{1}{l}{} & \multicolumn{1}{l}{}         & \multicolumn{1}{c}{}                                          & \multicolumn{1}{c}{}                                                      & \multicolumn{1}{l}{}       & \multicolumn{1}{l}{}       & \multicolumn{1}{l}{}       
		\end{tabular}
			}
		\vspace{-1.8cm}
			\end{table*}
		
			Table~\ref{tab:relatedwork} identifies and compares the main elements of related works with ours in terms of their IoT application, architectural, and application placement engine properties. In the IoT application section, the dependency model of each proposal is studied, which can be either independent or dependent. Moreover, we study how each proposal models IoT applications in terms of the number of tasks and heterogeneity. This demonstrates whether IoT applications consist of homogeneous or heterogeneous tasks in terms of their computation and data flow. In the architectural properties, the attributes of IoT devices, fog/edge servers, and cloud servers are studied. For IoT devices, the overall number of devices and their type of requests are identified. The heterogeneous request type shows that each device has a different number of requests with various requirements compared to other IoT devices. For edge/fog servers, the number of deployed servers between IoT devices and cloud servers and the heterogeneity of their resources are studied. Moreover, the multi-cloud shows either these works consider different cloud service providers with heterogeneous resources or not. In the application placement engine, the main employed DRL methods are identified. Besides, it is studied either these works consider any mechanism to provide priority for the execution of tasks or not. Finally, the decision parameters of these DRL-based techniques are identified. 
			\par
			Considering DRL-based application placement techniques in edge and fog computing and their identified properties, the environment with multiple heterogeneous IoT devices, heterogeneous FSs, and heterogeneous multi CSs has the highest number of features. Moreover, DAG-based IoT applications incur more constraints on DRL agents as they need to consider the dependency among tasks within each IoT application. The exploration cost of DRL agents increases as the number of features and complexity of the environment increases. It negatively affects the training and convergence time of DRL techniques, and accordingly users' experience. To address these issues, we propose a distributed DRL technique based on the IMPALA architecture, called $X$-DDRL, in which several actors independently interact with fog computing environments and create experience trajectories in parallel. Then, these distributed experience trajectories are forwarded to the learner for training and policy updates. This significantly reduces the exploration and training costs of centralized DRL techniques. Furthermore, since the learner directly uses the batches of experience trajectories of distributed actors, rather than gradients with respect to the parameters of the policy (similar to how the A3C algorithm works), it can more efficiently learn and identify the features of input data \cite{espeholt2018impala}. Also, the transmission of gradients among actors and learners is more expensive in terms of data exchange size and time (similar to how A3C works) in comparison to sharing trajectories of experience. Hence, experience-sharing DRL techniques such as IMPALA are more practical and data-efficient in highly distributed and stochastic environments \cite{espeholt2018impala}, such as fog computing. Since the policy used to generate the trajectories of experiences in distributed actors can lag behind the policy of the learner in the time of gradient calculations, a V-trace off-policy actor-critic algorithm is used to correct this discrepancy. Besides, to capture the temporal behavior of input data, we embed RNN layers in the network of actors and learners. Moreover, $X$-DDRL uses a replay buffer to improve the sample efficiency for training.
			
			\vspace{-1mm}
			\section{System Model and Problem Formulation}
			\label{system}
			
\begin{figure}[!t]
				\centering 
				%\captionsetup{justification=centering,margin=2cm}
				%	\hspace{0cm}
				\includegraphics[width=\linewidth,height=6cm, trim=0in 0in 0in 0in]{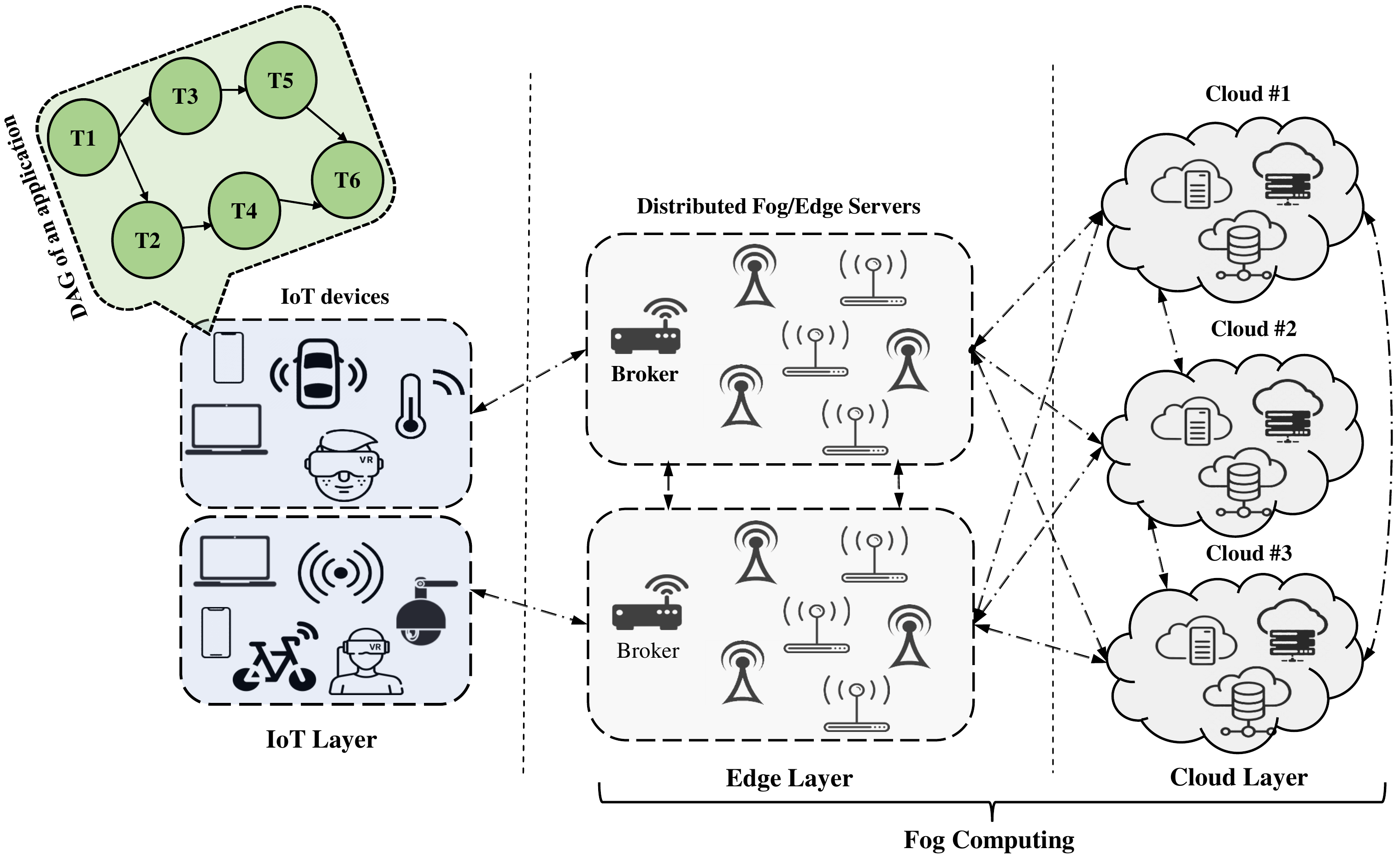}
				\caption{An overview of our system model}
				\label{fig:systemmodel}
				\vspace{-.6cm}
			\end{figure}

			Figure \ref{fig:systemmodel} represents an overview of our system model in fog computing. IoT devices send their application placement requests to brokers, situated at the edge of the network to be accessed with less latency and higher bandwidth \cite{mahmud2018fog,goudarzi2020application}. For each arriving application request, the broker makes a placement decision based on the corresponding DAG of the IoT application, its constraints, and the system status. Accordingly, each task of an IoT application may be assigned to the IoT device for the local execution or one of heterogeneous FSs or CSs for the execution. 
			
			%\vspace*{-.3cm}
			\subsection{IoT Application}
			
			Each IoT applications is modeled as a DAG $G=(\mathcal{V},\mathcal{E})$ of its tasks, where $\mathcal{V}=\{v_{i}|1\leq i \leq |\mathcal{V}|\}, |\mathcal{V}|=L$ depicts vertex set of one application, in which $v_i$ denotes the $i$th task. Moreover, $\mathcal{E}=\{e_{i,j}|v_{i}, v_{j} \in \mathcal{V},\thinspace i \neq j\}$ represents edge set, in which $e_{i,j}$ denotes there is a data flow between $v_i$ (i.e., parent), $v_j$ (i.e., child) and hence, $v_j$ cannot be executed before $v_i$. Accordingly, for each task $v_j$, a predecessor task set $\mathcal{P}(v_{j})$ is defined, containing all tasks that should be executed before $v_j$.  Moreover, for each DAG $G$, exit tasks are referred to tasks without any children.  			
			\par
			The amount of CPU cycles, required for the processing of each task, is represented as $v_{j}^{w}$, while the required amount of RAM for processing of each task is $v_{j}^{ram}$. Moreover, the weight on each edge $e_{i,j}^{w}$ illustrates the amount of data that task $v_i$ sends as its output to task $v_{j}$ as its input.  			
			\subsection{Problem Formulation}
			\label{sec:problem_formula}
			Each task of an IoT application can either be executed locally on the IoT device or on one of the FSs or CSs. We define the set of all available servers as $\mathcal{M}$ where $|\mathcal{M}|=M$. Each server is represented as $m^{y,z} \in \mathcal{M}$ where $y$ shows the type of server (IoT device ($y=0$), FSs ($y=1$), CSs ($y=2$)) and $z$ denotes the server index. Therefore, the placement configuration of task $v_j$, belonging to an IoT application, can be defined as:
			\begin{equation}
			x_{v_j}=m^{y,z}
			\end{equation}
			and accordingly, the placement configuration of an IoT application $\mathcal{X}$ is defined as the set of placement configurations for all of its tasks:
			\begin{equation}
				\mathcal{X}=\{x_{v_j}|v_j \in \mathcal{V},1\leq j \leq |\mathcal{V}|\}			\end{equation}	
			
			We consider that tasks of an IoT application are sorted in a sequence so that all parent tasks are scheduled for the execution before their children. Hence, the dependencies among tasks are satisfied. Besides, among tasks that can be executed in parallel (i.e., tasks that all of their dependencies are satisfied), the $CP(v_i)$ is an indicator function to demonstrate whether the task is on the critical path of the IoT application or not \cite{qi2019knowledge} (i.e., a path containing vertices and edges that incurs the highest execution cost). 
			
%			
%			\begin{eqnarray}
%				x_{i}= \Bigg \{ \begin{tabular}{cccc}0, & $s^{y,z}=s^{0}$, \\
%					1, & $s^{y,z} \in \{s^{1,1},s^{1,2},\cdots,s^{1,f}\}$ & $|z|=f$\\
%					2,&  $s^{y,z} \in \{s^{2,1},s^{2,2},\cdots,s^{2,c}\}$,&$|z|=c$
%				\end{tabular}
%				\label{criteria}
%			\end{eqnarray}
%		
%			where $x_{i}=0$ demonstrates that task $v_i$ is assigned to the IoT device ($s^{0}$), and $x_{i}=1$ and $x_{i}=2$ shows that task $v_i$ is assigned to one of FSs or CSs for remote execution, respectively. Moreover, the $f$ and $c$ show the number of available fog servers and cloud servers respectively.  
%			
%			
%			
%		
			
%			
%			We formulate the task placement problem as an optimization problem aiming at minimizing the overall execution time of IoT applications and energy consumption of IoT devices.
		\subsubsection{Execution time model}
%		Considering the Eq.~\ref{control}, the weighted cost optimization is equal to the execution time model when $\psi_{\gamma}=1$ and $\psi_{\theta}=0$.
		The execution time of each task $v_j$ depends on the availability time of required input data for that task $\psi_{x_{v_j}}^{input}$ and its processing time on the assigned server $\psi_{x_{v_j}}^{proc}$:
		
		\begin{equation} 
			\psi_{x_{v_j}}= \psi_{x_{v_j}}^{proc} + \psi_{x_{v_j}}^{input} 
		\end{equation}
		
		\noindent
		where $\psi_{x_{v_j}}^{proc}$ depends on the required CPU cycles for that task $v_{j}^{w}$ and the processing speed of the corresponding assigned server $f^{s}_{x_{v_j}}$, calculated as follows:
		
		\begin{equation}
			\psi_{x_{v_j}}^{proc}= \frac{v_{j}^{w}}{f^{s}_{x_{v_j}}}
		\end{equation}
		
		\noindent
		 The $\psi_{x_{v_j}}^{input}$ is calculated as the maximum time that the required input data for the execution of task $v_j$ become available on the corresponding assigned server (i.e., $x_{v_j}$) from its parent tasks:

%		\begingroup
%		\footnotesize	
%		\begin{equation}
%			\hspace{-3mm}
%		  	\psi_{x_{v_j}}^{input}= \max((\frac{e_{i,j}^{w}}{b_{x_{v_i},x_{v_j}}}+l_{x_{v_i},x_{v_j}})\times SS(x_{v_i},x_{v_j})),\;\;\;\; \forall v_i \in \mathcal{P}(v_j)
%		\end{equation}
%	     \endgroup 
	     
	     \begin{eqnarray}
	     	\psi_{x_{v_j}}^{input}= \max((\frac{e_{i,j}^{w}}{b_{x_{v_i},x_{v_j}}}+l_{x_{v_i},x_{v_j}})\times SS(x_{v_i},x_{v_j})),\\
	     	\forall v_i \in \mathcal{P}(v_j)\;\;\;\;\;\;\;\;\;\;\;\;\;\;\;\;\;\;\;\;\;\;\;\;\;\;\;\;\nonumber
	    \end{eqnarray}
		
		\noindent
		where $b_{x_{v_i},x_{v_j}}$ shows the current bandwidth (i.e., data rate) between the servers to which $v_i$ and $v_j$ are assigned, respectively. Moreover, $l_{x_{v_i},x_{v_j}}$ demonstrates the communication latency between two servers. The $SS(x_{v_i},x_{v_j})$ is equal to $0$ if $x_{v_i} = x_{v_j}$ (i.e., same assigned servers) or 1, otherwise. Since fog computing environments are heterogeneous and stochastic, the $f^{s}_{x_{v_j}}$, $b_{x_{v_i},x_{v_j}}$, and $l_{x_{v_i},x_{v_j}}$ may be different among IoT devices, FSs, and CSs. 
		\par		
		The main goal of the execution time model is to find the best-possible placement configuration for the IoT application so that its execution time becomes minimized. Assuming an IoT application consists of $L$ tasks, the execution time model is defined as:
		\begin{equation}
			\Psi(\mathcal{X}) = \min (\sum\limits_{j=1}^{L}CP(v_{j})\times \psi_{x_{v_j}})
			\label{timeModel}
		\end{equation}

		\noindent
		where $CP(v_j)$ is $1$ if task $v_j$ is on the critical path and $0$ otherwise. Due to the parallel execution of some tasks, only the execution time of tasks on the critical path is considered, which incurs the highest execution time and involves the execution time of other parallel tasks as well.

		\subsubsection{Energy consumption model}
		We only consider the energy consumption of IoT devices in this work since FSs and CSs are usually connected to constant power supplies \cite{xu2019computation}. From the IoT devices' perspective, the energy consumption that execution of each task $v_j$ incurs depends on the amount of energy the IoT device consumes until the required input data for that task $\omega_{x_{v_j}}^{input}$ becomes ready plus the required energy for the processing of that task $\omega_{x_{v_j}}^{proc}$:
		
		\begin{equation} 
			\omega_{x_{v_j}}= \omega_{x_{v_j}}^{proc} + \omega_{x_{v_j}}^{input} 
		\end{equation}
		
		\noindent
		where $\omega_{x_{v_j}}^{proc}$ depends whether the task is assigned to the IoT device for local execution or not. Hence, we define an IoT Server identifier $IS(x_{v_j})$ to show whether the $x_{v_j}$ refers to an IoT device ($IS(x_{v_j})=1$) or other servers ($IS(x_{v_j})=0$). Accordingly, the $\omega_{x_{v_j}}^{proc}$ is calculated as what follows:
		
		\begin{eqnarray}
			\omega_{x_{v_j}}^{proc}= \Bigg \{ \begin{tabular}{cccc}
				$\psi_{x_{v_j}}^{proc} \times P^{cpu}$, & $IS(x_{v_j})=1$ \\
				$\psi_{x_{v_j}}^{proc} \times P^{idle}$, & $IS(x_{v_j})=0$ 
			\end{tabular}
			\label{criteria}
		\end{eqnarray}
		
		\noindent
		If the task is assigned to the IoT device (i.e., $IS(x_{v_j})=1$), the energy consumption of the IoT device is equal to the amount of time that it processes the task multiplied by the CPU power of IoT device $P^{cpu}$. However, if the task is assigned to the other servers for processing (i.e., $IS(x_{v_j})=0$), the energy consumption of the IoT device depends on its idle time and corresponding idle power $P^{idle}$. 
		
		\noindent
		The $\omega_{x_{v_j}}^{input}$ depends on the assigned servers to current task (i.e., $x_{v_j}$) and its predecessors, and is calculated as what follows:
		
		\begingroup
		\footnotesize
		\begin{eqnarray}
			\label{eq:energyInput}
			\hspace{-2mm}
			\omega_{x_{v_j}}^{input}= \left \{ \begin{array}{ll}
				\psi_{x_{v_j}}^{input} \times P^{tra}, &IS(x_{v_j})=1
				\vspace{5mm}\\
				\max(IS(x_{v_i})\times (\frac{e_{i,j}^{w}}{b_{x_{v_i},x_{v_j}}}+l_{x_{v_i},x_{v_j}}) &IS(x_{v_j})=0\\ \times SS(x_{v_i},x_{v_j}))\times P^{tra}+(\psi^{idle} \times P^{idle}), &\\
				\forall v_i \in \mathcal{P}(v_j),&
			\end{array}\right.
			\label{criteria2}
		\end{eqnarray}
		\endgroup
		
		\noindent
		where $IS(x_{v_j})$ and $IS(x_{v_i})$ demonstrates whether the current task $v_j$ and/or its parent task $v_i \in \mathcal{P}(v_j)$ in each edge are assigned to the IoT device or not, respectively. It is important to note that the transmission energy consumption for each edge in DAG is only considered when one of the tasks is placed on the IoT device. Hence, if the current task is assigned to the IoT device (i.e., $IS(x_{v_j})=1$), the $\omega_{x_{v_j}}^{input}$ depends on the $\psi_{x_{v_j}}^{input}$. However, if the current task is not assigned to the IoT device (i.e., $IS(x_{v_j})=0$), it is possible that the predecessor tasks of the current task (i.e., $\forall v_i \in \mathcal{P}(v_j)$) are previously assigned to the IoT device, and hence the IoT device should forward the data to the server on which the current task is assigned (which incurs energy consumption). If none of the tasks are assigned to the IoT device for local execution, the IoT device is in its idle state. Besides, $P^{tra}$, $\psi^{idle}$ represent the transmission power of the IoT device and its idle time, respectively. Similar to \cite{xu2019computation,mahmoodi2016optimal,wu2019efficient}, we used constant values for $P^{tra}$, $\psi^{idle}$, however, these parameters also can be dynamically configured.	
		\par		
		The main goal of the energy consumption model is to find the best-possible placement configuration for the IoT application so that its energy consumption becomes minimized. Assuming an IoT application consists of $L$ tasks, the energy consumption model is defined as:
		\begin{equation}
			\Omega(\mathcal{X}) = \min (\sum\limits_{j=1}^{L}CP(v_j)\times \omega_{x_{v_j}})
			\label{energyModel}
		\end{equation}
					
		\subsubsection{Weighted cost model}
		The weighted execution cost of task $v_j$ is defined based on its assigned server $x_{v_j}$:
		 
		\begin{equation}
			\phi_{x_{v_j}}= (w_1 \times \psi_{x_{v_j}}) + (w_2 \times \omega_{x_{v_j}})
			\label{weightedTask}
		\end{equation}
		
		\noindent
		where $\psi_{x_{v_j}}$ and $\omega_{x_{v_j}}$ refer to the execution time and energy consumption for the execution of task $v_j$. Moreover, the $w_1$ and $w_2$ are control parameters to represent the importance of decision parameters in weighted execution cost of each task. Also, the weighted cost of each task can be changed to execution time or energy consumption cost of each task by assigning $w_1=1, w_2=0$ or $w_1=0, w_2=1$, respectively.
		\par
		Finally, the goal of weighted cost model is to find the best placement configuration for tasks of an IoT application while minimizing the weighted cost of parameters. In this work, we consider execution time of IoT applications and energy consumption of IoT devices as decision parameters, however, this weighted cost can be extended using other decision parameters. The weighted cost model is defined as:
		
%		\begin{equation}
%		\min \Phi(\mathcal{X}) = \min\limits_{w_1, w_2 \in [0,1]} w_1 \times \Psi(\mathcal{X}) + w_2 \times \Omega(\mathcal{X})
%		\label{TotalWeightedCost}
%		\end{equation}
\begin{equation}
	\min \Phi(\mathcal{X}) = \min w_1 \times \Psi(\mathcal{X}) + w_2 \times \Omega(\mathcal{X})
	\label{TotalWeightedCost}
\end{equation}
		$s.t.$
		\vspace{-0.3cm}
		\begin{eqnarray}
		&&C1:\;Size(x_{v_j})=1,\; \forall x_{v_j} \in \mathcal{X}\\
		&&C2:\;\Phi(v_i) \leq \Phi(v_i+ v_j),\; \forall v_{i}\in\mathcal{P}(v_{j})\\
		&&C3:\; v_j^{ram} \leq RAM (x_{v_j}), \; \forall v_j \in \mathcal{V}\\
		&&C4:w_1+w_2=1
		\end{eqnarray}
		\noindent
		where $\Psi(\mathcal{X})$, $\Omega(\mathcal{X})$ are obtained from Eq.~\ref{timeModel} and Eq.~\ref{energyModel}, respectively. Besides, $w_1$ and $w_2$ are control parameters for execution time and energy consumption, by which the weighted cost model can be tuned. $C1$ denotes that each task can only be assigned to one server at a time for processing. Moreover, $C2$ states that the task $v_j$ can only be executed after the execution of its predecessors, and hence the cumulative execution cost of $v_j$ is always larger or equal to execution cost of its predecessors' tasks \cite{xu2019computation}. Besides, $C3$ states that the assigned server to the task $v_j$ should have sufficient amount of available RAM $RAM(x_{v_j})$ for the processing. Also, $C4$ defines a constraint on the values of control parameters. These constraints are also valid for execution time and energy consumption models. Moreover, the weighted cost model can be changed to execution time or energy consumption model by assigning $w_1=1, w_2=0$ or $w_1=0, w_2=1$, respectively.
		\par
		Since the application placement problem in heterogeneous environments is an NP-hard problem \cite{qiu2020distributed}, the problem's complexity grows exponentially as the number of heterogeneous servers and/or the number of tasks within an IoT application increases. Thus, the optimal policy of the application placement problem cannot be obtained in polynomial time by iterative approaches. The existing application placement techniques are mostly based on heuristics, rule-based policies, and approximation algorithms \cite{wang2020fast,tuli2020dynamic}. Such techniques work well in general cases, however, they cannot fully adapt to dynamic computing environments where the effective parameters of workloads and computational resources continuously change \cite{wang2020fast,fox2019learning}. To address these issues, DRL-based scheduling/placement algorithms are promising avenues for dynamic optimizations of the system \cite{jeff2018ml,tuli2020dynamic}.
\begin{figure*}[t]
	%\begin{minipage}{\linewidth}
	\begin{subfigure}{.49\textwidth}
		\centering
		\includegraphics[height=6cm]{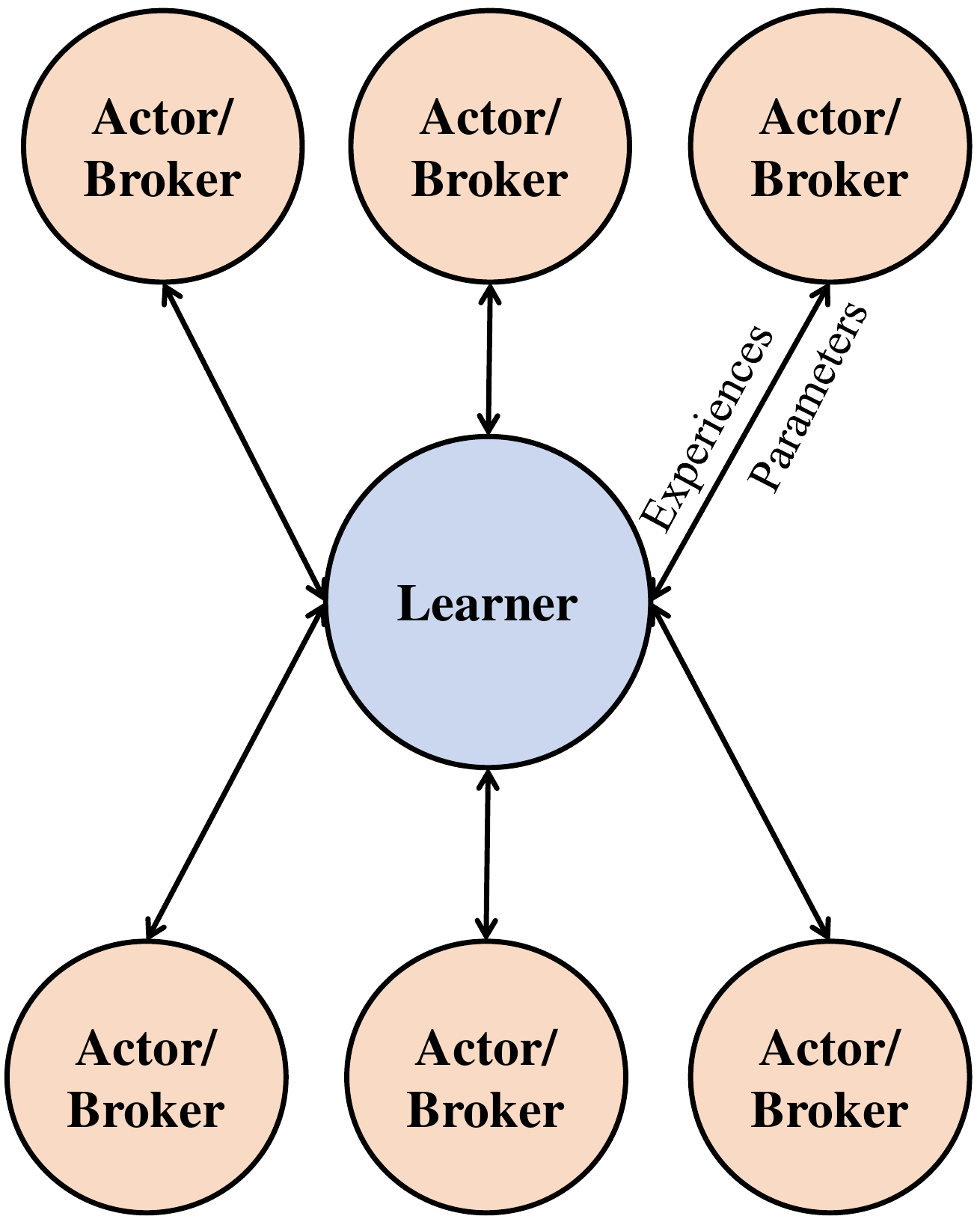}
		\captionsetup{justification=centering}
		\caption{High-level overview of learner and actors/brokers}
		\label{fig:actor_learner:1}
	\end{subfigure}%
	%\hspace{0.1cm}
	\begin{subfigure}{.49\textwidth}
		%\vspace{0.35cm}
		\centering
		\includegraphics[width=\linewidth,height=6cm]{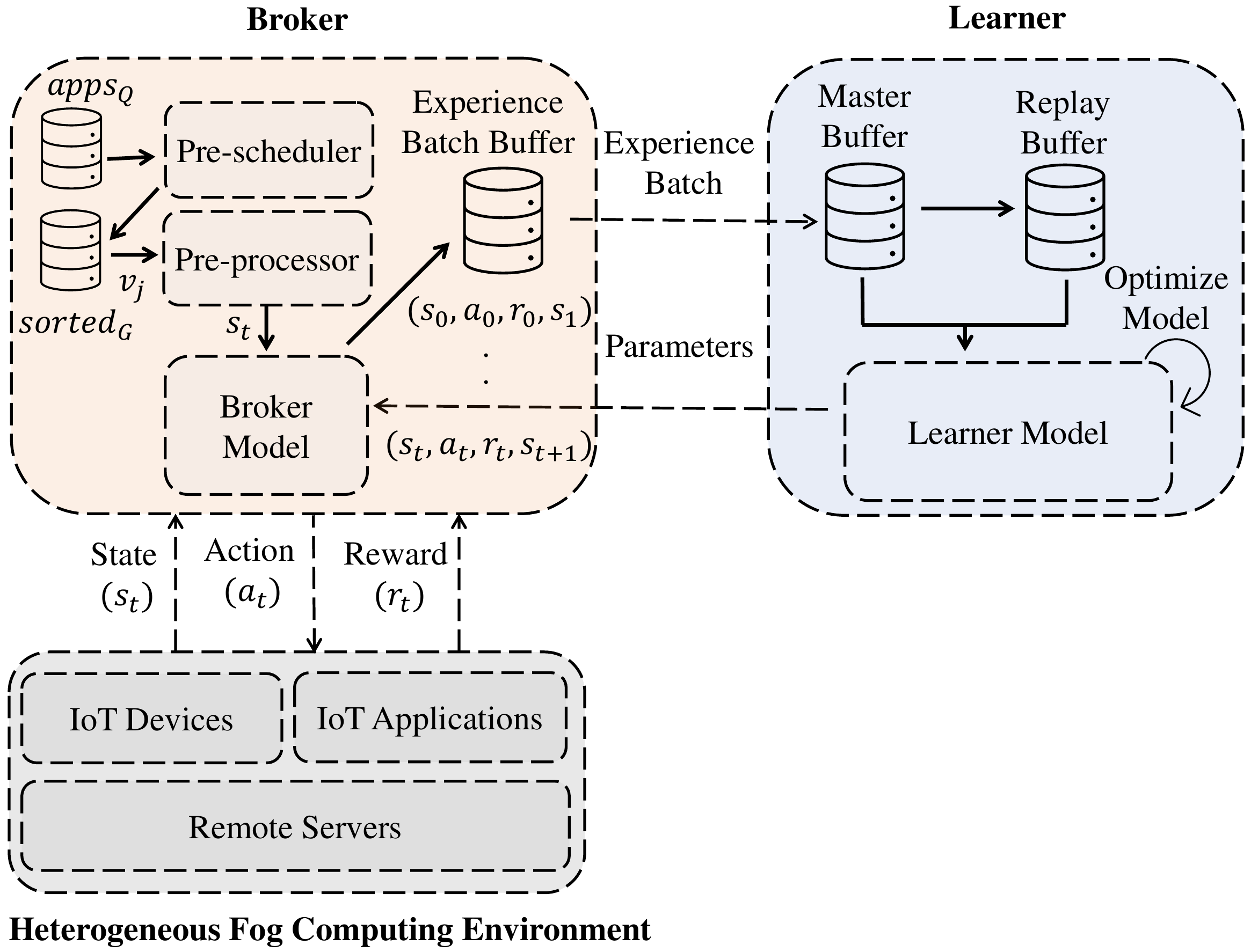}
		%\captionsetup{options}
		\captionsetup{justification=centering}
		\caption{Communication between learner and each broker}
		\label{fig:actor_learner:2}
	\end{subfigure}
	
	\caption{An overview of $X$-DDRL framework}
	\label{fig:actor_learner}
	\vspace{-0.5cm}
\end{figure*}  
			\section{Deep Reinforcement Learning Model}
			\label{sec:DRLModel}
			The DRL is a general framework that incorporates deep learning to solve decision-making problems with high-dimensional inputs. Formally, learning problems in DRL can be modeled as Markov Decision Processes (MDP), which is extensively used in sequential stochastic decision making problems. A learning problem can be defined by a tuple $<\mathbb{S},\mathbb{A},\mathbb{P},\mathbb{R},\gamma>$, in which $\mathbb{S}$ and $\mathbb{A}$ denote the state and action spaces, respectively. $\mathbb{P}$ illustrates the state transition probability, and $\mathbb{R}$ is a reward function. Finally, $\gamma \in [0,1]$ is a discount factor, determining the importance of future rewards. We suppose that the time horizon is separated into multiple time periods, called time steps $t \in \mathbb{T}$. The DRL agent interacts with the environment, and in each time step $t$, it perceives the current state of the environment $s_t$, and selects an action $a_t$ based on its policy $\pi(a_t|s_t)$, mapping states to actions. Considering the selected action $a_t$, the agent receives a reward $r_{t}$ from the environment, and it can perceive the next state $s_{t+1}$. The main goal of the agent is to find a policy in order to maximize the expected total of future discounted reward \cite{espeholt2018impala}:
			
			\begin{equation}
				\mathbb{V}^{\pi}(s_t) = \mathbb{E}_{\pi} [\sum\limits_{t \in \mathbb{T}}\gamma^{t}r_t]
			\end{equation}
			
			\noindent
			where $r_t = \mathbb{R}(s_t, a_t)$ is the reward at time step $t$, and $a_t \thicksim \pi(.|s_t)$ is the generated action at time step $t$ by following the policy $\pi$. Moreover, when DNN is used to approximate the function, the parameters are denoted as $\theta$.
			\par
			Considering the application placement in fog computing environments, we define the main concept of the DRL for our problem as what follows:
			\begin{itemize}
				\item \textbf{State space $\mathbb{S}$:} In our application placement problem, the state is the observations of the agent from the heterogeneous fog computing environment. Thus, the state at time step $t$ ($s_t$) consists of information about all heterogeneous servers (such as processing speed of CPU, number of CPU cores, CPU utilization, access Bandwidth (i.e., data rate) of servers, access latency of servers, and CPU, transmission, and idle power consumption values of IoT device). For the rest of the servers, their power consumption values are ignored as we only consider energy consumption from IoT devices' perspective \cite{xu2019computation}. If for each server we have $n$ features to represent its information, the feature vector of all $M$ servers at time step $t$ ($FV^{\mathcal{M}}_{t}$) can be presented as:
				\begin{equation}
					FV^{\mathcal{M}}_{t} = \{f_i^{m^{y,z}}|\forall m^{y,z} \in 	\mathcal{M},1\leq\ i \leq n\} 
					\label{FV_servers}
				\end{equation}
				
				\noindent
				where $f_i^{m^{y,z}}$ shows the $i$th feature of the server $m^{y,z}$. Moreover, $s_t$ contains the information about the current task to be processed within a DAG of an IoT application (such as computation requirements of the task, required RAM, amount of output data per parent task, and current placement configuration of all tasks). Since we consider that tasks are sorted and their dependencies are satisfied before their execution, the current placement configuration of tasks contains the information regarding assigned servers to all previous tasks. The values of unprocessed tasks are set to $-1$. If we assume that each task has $b$ features, the feature vector of task $v_j$ ($FV^{v_j}_{t}$) can be represented as:
				\begin{equation}
					FV^{v_j}_{t} = \{f_i^{v_j}| v_j \in \mathcal{V}, \forall i \; 1\leq i \leq b\} 
					\label{FV_tasks}
				\end{equation}		    
				
				\noindent
				where $f_i^{v_j}$ shows the $i$th feature of the task $v_j$. Thus, the system space can be defined as:
				\begin{equation}
					\mathbb{S}=\{s_t|s_t = (FV^{\mathcal{M}}_{t}, FV^{v_j}_{t}), \forall t \in \mathbb{T}\}
					\label{System_State}
				\end{equation}
				
				\item \textbf{Action space $\mathbb{A}$:} Actions are assignments of available servers to tasks of an IoT application. Therefore, the action at time step $t$ ($a_t$) is equal to assigning a server $m^{y,z}$ to the current task $v_j$. Considering the placement configuration of each task $x_{v,j}$ in Section~\ref{sec:problem_formula}, $a_t$ can be defined as:
				\begin{equation}
					a_t = x_{v_j}= m^{y,z}
				\end{equation}
				\noindent
				Thus, the action space $\mathbb{A}$ can be defined as the set of all available servers, presented as follows:
				 
				 \begin{equation}
				 	\mathbb{A} = \mathcal{M}
				 \end{equation}
				
%				\item \textbf{State transition probability $\mathbb{P}$:} It represents the probability of transition to a new state while considering the current state and action.  

				 \item \textbf{Reward function $\mathbb{R}$:} The goal is to minimize the weighted cost model, defined in Eq.~\ref{TotalWeightedCost}. To obtain this, we consider Eq.~\ref{weightedTask} as the weighted cost of each task and define the $\mathbb{R}$ as the negative value of Eq.~\ref{weightedTask} if the task can be executed ($done=1$). Moreover, we define a constant $penalty$ value, which is usually a very large negative number~\cite{qiu2020distributed}. Furthermore, the $penalty$ value can be dynamically set based on the goal of the optimization problem and environmental variables. If the selected action by the agent (i.e., server assignment for the current task) cannot be performed due to any reason ($done=0$), the reward becomes equal to $penalty$. Accordingly, $r_t$ is defined as:
				 
				 \begin{eqnarray}
				 	\label{eq.rewardFunctionTimeStep}
				 	r_t= \left\{ \begin{tabular}{cc} $-\phi_{x_{v_j}}$ & $done=1$ \vspace{0.1cm}\\	
				 					$penalty$ & $done=0$
				 	\end{tabular}\right.
				 \end{eqnarray}
				 
%				 \item \textbf{Agent:} It refers to the algorithm used to learn the optimal policy for the DAG-based application placement problem.			
				 
		\end{itemize}			
			\section{Distributed DRL-based Framework}
			\label{placement}
%			In this section, we briefly discuss main steps of DRL and how application placement in fog computing environment can be modeled accordingly. Then, we discuss our proposed distributed application placement technique in detail. 
			To address the challenges of DAG-based application placement in the heterogeneous fog computing environment, the $X$-DDRL works based on an actor-critic framework, aiming at taking advantage of both value-based and policy-based techniques while minimizing their drawbacks \cite{qi2019knowledge}. 
			
			\paragraph*{\textbf{Actor-critic framework}}
			In an actor-critic framework, the policy is directly parameterized, denoted as $\pi(a_t|s_t;\theta)$, and the $\theta$ is updated by calculating the gradient ascent on the variance of the expected total future discounted reward (i.e, $\sum\limits_{k=0}^{\infty}\gamma^{k}r_{t+k}$) and the learned state-value function under policy $\pi$ (i.e., $\mathbb{V}^{\pi}(s_t)$) \cite{qi2019knowledge}. The actor interacts with the environment and receives state $s_t$, outputs the action $a_t$ based on $\pi(a_t|s_t;\theta)$, and receives the reward $r_t$ and next state $s_{t+1}$. The critic, on the other hand, uses rewards to evaluate the current policy based on the Temporal Difference (TD) error between current reward and the estimation of the value function $\mathbb{V}(s_t;\theta)$. Both actor and critic use DNNs as their function approximators, which are trained separately. To improve the selection probability of better actions by the actor, the parameters of the actor network are updated using the feedback of the TD-error, while the network parameters of the critic network are updated to achieve better value estimation. While the actor-critic frameworks work very well in long-term performance optimizations, their learning speeds are slow and they incur huge exploration costs, especially in problems with high dimensional-state space \cite{qiu2020distributed}. The distributed learning techniques in which diverse trajectories are generated in parallel can greatly improve the exploration costs and learning speed of actor-critic frameworks.
			\par
			The $X$-DDRL works based on an actor-learner framework, in which the process of generating experience trajectories is separated from learning the parameters of $\pi$ and $\mathbb{V}^{\pi}$. Fig~\ref{fig:actor_learner:1} demonstrates a high-level overview of learner and actors. The distributed actors in fog computing environments, which can be multiple CPUs within a broker (i.e., FS) or different brokers, interact with their fog computing environments. Arriving application placement requests to each broker are queued in the $apps_{Q}$ based on the First-In-First-Out (FIFO) policy. As Fig~\ref{fig:actor_learner:2} depicts, brokers performs pre-scheduling phase for each IoT application. Then, based on features of available servers and current task of selected IoT application, each broker pre-processes the current state and makes an application placement decision. Each broker periodically sends its local experience trajectories to the learner. Besides, the learner updates the target policy $\pi$ based on collection of received trajectories from different brokers and past trajectories stored in the replay buffer. After each policy update of the learner, brokers update their local policy $\mu$ with the policy of the learner $\pi$.
			\par
			The $X$-DDRL is divided into two phases: pre-scheduling and application placement technique. In the pre-scheduling, tasks of the received IoT application are ranked and sorted in a sequence for the execution. Afterward, for each task of an IoT application, $X$-DDRL makes a placement decision to minimize the execution cost of the IoT application.
			\subsection{$X$-DDRL: Pre-scheduling Phase}
			\label{sec:pre-scheduling}
\begin{figure}[!t]
				\centering 
				%\captionsetup{justification=centering,margin=2cm}
				%	\hspace{0cm}
				\includegraphics[height=5cm]{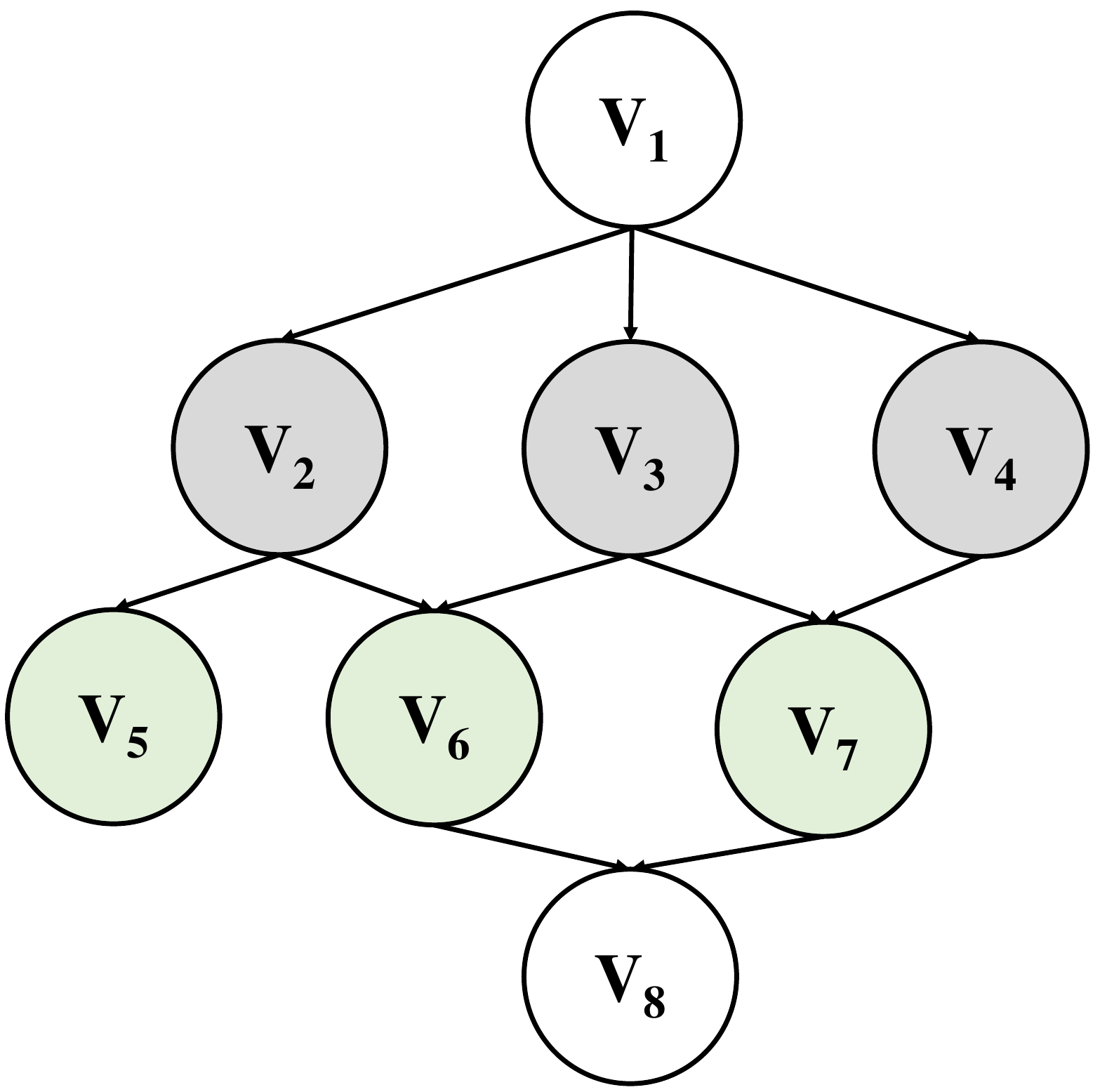}
				\caption{A sample IoT application (parallel tasks have same colors in each row)}
				\label{fig:IoTapplication}
				\vspace{-.1cm}
			\end{figure}

			IoT applications are heterogeneous in terms of the number of tasks per application, the dependency model, and corresponding weights of vertices and edges. Considering the dependency model of an IoT application, tasks should be sorted for execution, so that task $v_j$ cannot be executed before any task $v_i \in \mathcal{P}(v_{j})$. Furthermore, there are several tasks that can be executed in parallel, and the order of execution of such parallel tasks are also important and may affect the execution cost of an IoT application. Figure \ref{fig:IoTapplication} shows a sample IoT application, dependencies among tasks, and parallel tasks with the same colors in each row.
			\par
			Whenever a broker receives a DAG-based IoT application request from a user, it creates a sequence of tasks for the execution while considering above-mentioned challenges. Tasks within the IoT application are ranked based on the non-increasing order of their rank value. The rank value of a task is defined as: 
			
			\begingroup
			\footnotesize
			\begin{eqnarray}
				\label{eq.upwardRank}
				Rank(v_{j})= \left\{ \begin{tabular}{cc} $\widetilde{\phi_{x_{v_j}}} + \max(\widetilde{\phi_{x_{v_i}}})$ &\text{if } $v_{n,j} \neq exit$ \vspace{0.1cm}\\	
					$\forall v_{i}\in \mathcal{P}(v_{j})$,&\vspace{0.5cm}	\\
					$\widetilde{\phi_{x_{v_j}}}$, &\text{if } $v_{n,j} = exit$
				\end{tabular}\right.
			\end{eqnarray}
			\endgroup
			
			\noindent
			where $\widetilde{\phi_{x_{v_j}}}$ shows the average weighted execution cost of task $v_{n,j}$ on considering different servers. The rank is calculated recursively by traversing the DAG of the application, starting from the exit module. Using the rank function, the tasks on the critical path of DAG (i.e., $CP$) can also be identified. Hence, not only does the rank function satisfy the dependency among tasks, but it also defines an execution order for tasks that can be executed in parallel. To achieve this, it gives higher priority to tasks that incur higher total execution costs among parallel tasks. 

			\subsection{$X$-DDRL: Application Placement Phase}
			 
%			However, due to the distributed training, a policy gap may happen between the policy of actors and critics, and hence, the learning becomes off-policy.    
			%\subsubsection{$X$-DDRL algorithm process}
			\begin{algorithm}[!t]
				\footnotesize
				\caption{The role of each broker/actor} \label{alg:actor}
				\SetKwData{Left}{left}
				\SetKwData{This}{this}
				\SetKwData{Up}{up}
				\SetKwFunction{Union}{Union}
				\SetKwFunction{FindCompress}{FindCompress}
				\SetKwInOut{Input}{Input}
				\SetKwInOut{Output}{Output}
				\SetKwInOut{Parameter}{Parameter}
				\Input{$\pi$: The learner policy}
				%\Output{$finalConfigs$, $finalCost$}
				\tcc{$N$: number of steps, $\mu$: the actor's local policy, $EBB$: expeience batch buffer, $apps_{Q}$: Queue of all received IoT applications, $G$: current IoT application}
				% 	$MAResultList$ = $null$\\
				flag-init=True\\
				\For {$t=0$ to $\infty$}{
					$\mu$=UpdateLocalPolicy($\mu$, $\pi$)\\
					\For{$i=t$ to $N+t-1$}{
						\eIf{flag-init=True}{
							$G$=$apps_{Q}$.dequeue()\\
							$sorted_G$ = Pre-scheduling ($G$) \% based on Eq.~\ref{eq.upwardRank}\\
							$s_i$=ReceiveInitialState($G$, $\mathcal{M}$, $sorted_G$) \\
							flag-init=False\\
						}
						{$s_i$=ReceiveCurrentState() \\}
						$s_i$=Pre-processor($s_i$) \\
						$a_i$=PlacementEngine($s_i$, $\mu$)  \% calculates the action\\
						\nonl\%The environment then executes this action  \\
						%$s_{n+1}$=SendTaskForProcessing($a_n$) \\ 
						$r_i$=TaskCostCalculator($s_i$, $a_i$)  \% baed on Eq.~\ref{eq.rewardFunctionTimeStep} \\
						$s_{i+1}$ = BuildNextState($s_{i}$, $a_i$)\\
						$EBB$.update($s_i$, $a_i$, $r_i$, $s_{i+1}$)\\
						\If{Finish($G$)}{
							CalculateTotalCost($G$) \% based on Eq.~\ref{TotalWeightedCost}\\ 
							flag-init=True\\
						}			
					}
					%\if
					%TrajectoryBuilder
					\If{size($EBB$)==$N$}{
						SendExpeienceToLearner($EBB$)\\ 
					}
					
				}
	
			\end{algorithm}
		If we assume that each broker makes placement decisions for tasks of IoT applications, using their local policy $\mu$, for $N$ steps in the time horizon starting at time $i=t$, algorithm \ref{alg:actor} shows how brokers perform application placement decisions and generate experience trajectories. Each broker performs the following steps: At the beginning of each trajectory, the broker updates its policy $\mu$ with the policy of the learner (line 3). When broker starts making placement decisions for tasks of a new IoT application $G$ (i.e., when the flag-init=$True$), it receives the current IoT application from the $apps_{Q}$ (contains all received application requests to this broker) (line 6). Then, the broker performs the pre-scheduling to obtain the sorted list of application' tasks based on the Eq.~\ref{eq.upwardRank} (line 7). Next, the system state is generated using the initial state of the IoT application $G$ and available servers $\mathcal{M}$ (line 8). Moreover, the broker changes the flag-init to $False$, indicating that in the subsequent steps there is no need to re-calculate the ranking and initial state of the $G$ (line 9), and the broker only requires to obtain the current state of the environment based on Eq.~\ref{System_State} (line 11). The current state of the broker's environment $s_i$ consists of feature vectors of servers $FV^{\mathcal{M}}_{t}$ and the current task of IoT application $FV^{v_j}_{t}$. The current task of each IoT application is obtained from the ordered sequence of tasks $sorted_{G}$. Then, the broker pre-processes and normalizes values of the current state (line 13). Considering $s_i$ and current policy $\mu$, an application placement decision (i.e., the assignment of a server for the processing of the current task) is made (line 14). The current task is then forwarded to the assigned server (based on $a_i$) for processing. After the execution of the task, the broker receives the reward of this action, which is the negative value of the weighted execution cost of this task Eq.~\ref{eq.rewardFunctionTimeStep} (line 15). The next state of the environment is then created using the $BuildNextState$ function (line 16). Then, the broker creates an experience tuple ($s_i$,$a_i$,$r_i$,$s_{i+1}$) and stores it in its local experience batch buffer (line 17). When the broker finishes assignment of servers to all tasks of the current IoT application $G$, meaning the application placement is done for the current IoT application, the total weighted execution cost of this IoT application is calculated using Eq.~\ref{TotalWeightedCost} (line 19). Moreover, the broker sets flag-init to $False$ so that the next IoT application in the queue of this broker $apps_{Q}$ can be served (line 20). After $N$ steps, each broker forwards its experience batch buffer to the learner (lines 23-25). The learner periodically updates its policy (i.e., $\pi$) on batches of experience trajectories, collected from several brokers. 			
		\par
		Since policies of brokers $\mu$ are updated based on the learner's policy (trained on trajectories of different brokers), each broker gets the benefit of trajectories generated by other brokers. It significantly reduces the exploration cost of each broker, and also provides brokers with a more accurate local policy $\mu$.  Furthermore, the $X$-DDRL uses an experience-sharing approach, which significantly reduces communication overhead between brokers and learners, in comparison to gradient-sharing techniques such as A3C \cite{espeholt2018impala}. 
		\par
		Due to the gap between the policy of broker $\mu$ (when generating new decisions) and the policy of the learner $\pi$ in the training time (when the learner estimates the gradients), the learner in the $X$-DDRL uses the off-policy correction method, called V-trace \cite{espeholt2018impala}, to correct this discrepancy.
		\begin{algorithm}[!t]
			\footnotesize
			\caption{The role of each learner} \label{alg:learner}
			\SetKwData{Left}{left}
			\SetKwData{This}{this}
			\SetKwData{Up}{up}
			\SetKwFunction{Union}{Union}
			\SetKwFunction{FindCompress}{FindCompress}
			\SetKwInOut{Input}{Input}
			\SetKwInOut{Output}{Output}
			\SetKwInOut{Parameter}{Parameter}
			\Input{$EB_{broker}$: Experience batch of different brokers}
			%\Output{$finalConfigs$, $finalCost$}
			\tcc{$list_{brokers}$: list of brokers, $\pi$: the learner's policy, $MB$: master buffer, $MBS$: master buffer size, $RB$: replay buffer, $RBS$: replay buffer size, $TB$: training batch, $TBS$: training batch size}
			\While {True}{
				flag-training=False\\
				%	$\mu$=UpdateLocalPolicy($\mu$, $\pi$)\\
				$MB$=$\emptyset$\\
				\While{flag-training==False}{
					$MB$.update($EB_{broker}$)\\
					\If{$TBS\leq MBS+RBS$}{
						TB=BuildTrainBatch($MB$, $RB$)\\
						flag-training==True
					}
				}
				OptimizeModel($TB$) \% based on Eq.~\ref{eq:vtraceupdate:1}, \ref{eq:vtraceupdate:2}\\
				UpdateBrokers($list_{brokers}$, $\pi$)	\\		
			}

		\end{algorithm}
			 
			 \begin{itemize}
			\item{\textbf{V-trace off-policy correction method}}: We assume that each broker generates an experience trajectory for $N$ steps while following its local policy $\mu$ as $(s_t,a_t,r_t)_{t=i}^{i+N}$. The value approximation of state $s_i$, defined as $N$-step V-trace target for $\mathbb{V}(s_i)$, is as follows:
			 
			 \begin{equation}
			   	\overline{\mathbb{V}_i}=\mathbb{V}(s_i)+\sum\limits_{t=i}^{i+N-1} \gamma^{t-i}(\prod\limits_{j=i}^{t-1}c_j)\delta_t\mathbb{V}
			   	\label{eq:v-trace-n-step}
			 \end{equation}
			\noindent
			where $\delta_t\mathbb{V}$ is a TD for $\mathbb{V}$, defined as: 
			\begin{equation} 
			 \delta_t\mathbb{V} = \rho_t(r_t+\gamma \mathbb{V}(s_{t+1})-\mathbb{V}(s_t))
			\end{equation}    
			\noindent
			where $\rho_t=min(\overline{\rho},\frac{\pi(a_t|s_t)}{\mu(a_t|s_t)})$ and $c_j=min(\overline{c},\frac{\pi(a_j|s_j)}{\mu(a_j|s_j)})$ are truncated Importance Sampling (IS) weights, while $\overline{c} \leq \overline{\rho}$. The $\overline{c}$ and $\overline{\rho}$ play different roles in the V-trace. The $\overline{\rho}$ has a direct effect on the value function $\mathbb{V}^{\pi}$ toward which we converge, while $\overline{c}$ has a direct effect on speed of the convergence. Considering $\rho$, the target policy of the learner $\pi$ can be defined as:
			
			\begin{equation}
				\pi_{\overline{\rho}}(a|s)=\frac{\min(\overline{\rho}\mu(a|s),\pi(a|s))}{\sum_{b \in \mathbb{A}}\min(\overline{\rho}\mu(b|s),\pi(b|s)))}
			\end{equation}
		\end{itemize}
			\par 
			We consider: (1) the brokers generate trajectories while following policy $\mu$, (2) the parameterized state-value function under $\theta$ as $\mathbb{V}_{\theta}$, (3) the current policy of learner is $\pi_{u}$, and (4) the V-trace target $\overline{\mathbb{V}_i}$ is defined based on Eq.~\ref{eq:v-trace-n-step}. The learner updates value parameters $\theta$, at time step $i$, in the direction of:
			\begin{equation}
				\label{eq:vtraceupdate:1}
				(\overline{\mathbb{V}_i}-\mathbb{V}_{\theta}(s_i))\nabla_{\theta}\mathbb{V}_{\theta}(s_i)
			\end{equation}
			Moreover, the policy parameters $u$ are updated in the direction of the policy gradient using Adam optimization algorithm \cite{kingma2015adam}:
			\begin{equation}
				\label{eq:vtraceupdate:2}
				\rho_i\nabla_u\log(\pi_u(a_i|s_i))(r_i+\gamma\overline{\mathbb{V}_{i+1}}-\mathbb{V}_{\theta}(s_i))
			\end{equation}
		\par
		Algorithm~\ref{alg:learner} summarizes the learners' role in the $X$-DDRL. The learner continuously receives and stores experience batches of brokers $EB_{broker}$ and updates the master Buffer $MB$ until the training batch $TB$ becomes full (line 4-10). Then, the learner optimizes the current target policy $\pi$ based on Eq.~\ref{eq:vtraceupdate:1} and \ref{eq:vtraceupdate:2} (line 11). After policy update of the learner, brokers update their local policies $\mu$ with the latest policy of the learner $\pi$ (i.e., brokers set their policies to the new learner policy), and hence, new application placement decisions are made using the updated policy $\mu$ in the brokers. The learner in the $X$-DDRL uses the replay buffer $RB$, which remarkably improves sample efficiency. The $X$-DDRL can easily scale as the number of servers, IoT application requests, and brokers increases, which is a principal factor in highly distributed environments such as fog computing. If a new broker joins the environment, the broker updates its local policy with the latest policy of the learner, and hence it takes advantage of all trajectories that previously generated by other brokers. Besides, it generates new sets of trajectories which help to better diversify the trajectories of the learner. If the number of servers in the environment increases, distributed brokers quickly generate new sets of trajectories, and accordingly the learner can update its target policy promptly. Such a collaborative distributed broker-learner architecture not only significantly improves the exploration costs, but also improves the convergence speed. The other improvement in the $X$-DDRL is using RNN layers since they can accurately identify highly non-linear patterns among different input features, resulting in significant speedup in the learner \cite{appleyard2016optimizing,tuli2020dynamic}.
        
        \subsection{Discussion on Resource Contention}		
		In heterogeneous computing environments where multiple applications are forwarded to heterogeneous servers, resource contention for computing resources may occur. Let's assume, there are three IoT applications named $A1$, $A2$, and $A3$ while there are two servers (either at the edge or cloud) called $S1$ and $S2$. The type of applications may be different from each other with different resource requirements, and the number of servers or their computing capability may differ so that resource contention occurs among IoT applications. Moreover, for the DAG-based IoT applications, consisting of several dependent tasks, another type of resource contention may happen among tasks of one IoT application.
\par 
One approach to solve the resource contention, either among different IoT applications or tasks of one application, is prioritization. In $X$-DDRL, the FIFO policy is used to prioritize different IoT applications. These policies can be changed according to the targeted problem. Besides, for the tasks of one application, there are two important points to consider. In DAG-oriented IoT application, each task within the application can only be executed if its predecessor tasks are completed. However, for tasks that can be executed in parallel, a priority should be defined. In $X$-DDRL, we use a rank function, Eq.~\ref{eq.upwardRank}, that prioritizes tasks of one IoT application while considering the dependency among tasks and the average execution cost of tasks. Such prioritization between different IoT applications and tasks of one IoT application help to solve the resource contention.
\par
Moreover, the DRL agent receives the sequence of tasks based on the above-mentioned policies. So, these prioritization techniques are very important for the long-term reward, especially for DAG-based IoT applications having extra constraints. Without such prioritization, the DRL agent may converge to a good solution but the convergence time is significantly higher while in some scenarios the DRL agent even cannot converge to good solution. The DRL agent learns to assign the best server to each task while considering the tasks' dependency model of an IoT application, resource requirements of each task, and available heterogeneous resources in the environment. That is, the DRL agent is considering the resource contention while assigning a server to each task, to minimize the execution cost of each task (based on short-term reward) and accordingly each IoT application (based on long-term reward).         			
			\section{Performance Evaluation}
			\label{evaluation}
			This section first describes the experimental setup, used to evaluate our technique and baseline algorithms. Next, the hyperparameters of our proposed technique $X$-DDR are discussed. Finally, we study the performance of $X$-DDRL and its counterparts in detail.
			
			\subsection{Experimental Setup}
			To evaluate the performance of the $X$-DDRL, we use both simulation environment and testbed, which their specification are provided in what follows.
						
			\subsubsection{Simulation setup}
			\label{sec:simulationSetup} 
			We developed an event-driven simulation environment in Python using the OpenAI Gym \cite{brockman2016openai} for the application placement in heterogeneous fog computing environments, similar to \cite{wang2020fast}. For each of the two learners, we set the number of brokers to $8$, which have access to a set of servers, and make application placement decisions accordingly. Hence, we vectorized the fog computing environment, generated using OpenAI Gym, so that distributed brokers can interact with their fog computing environments and make application placement decisions in parallel.
			Unlike prior work \cite{wang2020fast,qiu2020distributed,gazori2020saving,lu2020optimization}, we consider a heterogeneous fog computing environment consisting of IoT devices, resource-constrained FSs, and resource-rich CSs. In fog computing environment, we used the following server setup, unless it is stated in the experiments: two Raspberrypi 3B (Broadcom BCM2837 4 cores @1.2GHz, 1GB RAM)\footnote{https://www.raspberrypi.org/products/raspberry-pi-3-model-b}, one Raspberrypi 4B (ARM Cortex-A72 4 cores @1.5GHz, 4GB RAM)\footnote{https://www.raspberrypi.org/products/raspberry-pi-4-model-b}, and one Jetson Nano (ARM Cortex-A57 4 cores @1.43GHz, 4GB RAM, 128-core Maxwell GPU)\footnote{https://developer.nvidia.com/embedded/jetson-nano-developer-kit} as heterogeneous FSs. Besides, to simulate a heterogeneous multi-cloud environment, we used specifications of six m3.large instances of Nectar Cloud infrastructure (AMD 8 cores @2GHz, 16GB RAM)\footnote{https://nectar.org.au/} and two instances of the University of Melbourne Horizon Cloud (Intel Xeon 8 cores @2.4GHz, 24GB RAM, NVIDIA P40 3GB RAM GPU)\footnote{https://people.eng.unimelb.edu.au/lucasjb/horizon/}. For IoT devices, the server type is a single core @1GHz device embedded with 512MB RAM \cite{wang2020fast}. Besides, the power consumption of each IoT device in processing, idle, and transmission state is 0.5W, 0.002W, and 0.2W, respectively \cite{xu2019computation}. The bandwidth (i.e., data rate) and latency among different servers and IoT devices are also obtained based on average profiled values from testbed, similar to \cite{tuli2020dynamic}. Hence, the latency of FSs and CSs are considered as 1ms and 10ms respectively, similar to \cite{tuli2020dynamic}. The bandwidth between IoT devices and FSs is randomly selected between 10-12MB/s, while the bandwidth between IoT devices and FSs to the CSs is randomly selected between 4-8 MB/s, similar to \cite{wu2019efficient}. Although we obtained these values based on testbed experiments, they are referred to some similar works as well to show the credibility of these values. Also, both $w_{1}$ and $w_2$ are set to 0.5, meaning that the importance of execution time and the energy consumption is equal in the results. However, these parameters can be adjusted based on the users' requirements and network conditions.
			\par
		%	\vspace{-2mm}
			Many real-world IoT applications can be modeled by DAGs with a different number of tasks and dependency models. Hence, we generated several synthetic DAG sets with a different number of tasks and dependency models to represent scenarios where IoT devices generate heterogeneous DAGs with different preferences, similar to \cite{arabnejad2013list, wang2020fast}. The dependency model of each DAG can be identified using three parameters: number of tasks within an application $L$, $fat$ that controls the width and heights of a DAG, and $density$ that identifies the number of edges between different levels of the DAG. Accordingly, we generated different DAG datasets, where each dataset contains 100 DAGs with a similar number of tasks, fat, and density while the weights are randomly selected to represent heterogeneous task requirements in IoT applications with the same DAG structure. To generate heterogeneous DAG datasets, we set task numbers $L \in \{10,15,20,25,30,35,40,45,50\}$, $fat \in \{0.4,0.5,0.6,0.7,0.8\}$, and $density \in \{0.4,0.5,0.6,0.7,0.8\}$. To illustrate, one dataset of DAGs is $L=10$, $fat=0.4$, and $density=0.4$, containing 100 DAGs. Accordingly, for each task number $L$, we have 25 different combinations of $fat$ and $density$, resulting in 25 different topologies and 2500 DAGs. Finally, the simulation experiments are all performed on an instance of Horizon Cloud with the above-mentioned specifications.
			%			 Table \ref{tab:serverparameters} provide a summary of parameters in the simulation setup, consisting of number of servers in the environment and specifications of each server. 
		
			\subsubsection{Testbed setup}
			\label{sec:real-testbed}
			To evaluate the performance of $X$-DDRL in a real-world scenario, we created a testbed, similar to \cite{qiu2020distributed,hu2019learning}. The type of servers are the same as simulation setup while the number of servers of each type is as follows: two Raspberry pi 3B, one Raspberry pi 4B, one Jetson Nano, one instance of Horizon Cloud, and six m3.large instances of Nectar Cloud infrastructure). As IoT devices, we created several single-core VMs within a PC (HP Elitebook 840 G5 with Intel Core i7-8550U 8 cores @2GHz and 16GB RAM). These VMs are used to send application placement requests, using described DAG datasets, to the brokers. Moreover, to estimate the energy consumption of IoT devices, we used computing power, transmission power, and idle power as discussed in Section~\ref{sec:simulationSetup}, similar to the approach in \cite{qiu2020distributed}.
			For the connectivity, we set up a virtual network using VPN among IoT devices, FSs, and CSs, as described in \cite{deng2021fogbus2,goudarzi2021resource}. Due to the limited CPU and RAM of the IoT devices' VMs, they can send application placement requests, using a message-passing protocol (implemented using HTTP requests), to the broker that is the Jetson Nano in this testbed. The broker runs a multi-threaded server application that receives application placement requests from different IoT devices and puts them in the queue based on the FIFO policy. The broker dequeues the requests and makes placement decisions for the tasks according to its policy $\mu$. According to the placement configuration for each IoT application, each server that receives a task for processing assigns that task to one of its threads for processing. The thread is kept busy according to the weight of task and processing speed of the server. After the execution of each task, the size of output results that should be forwarded to the children tasks is obtained based on the weights of the task's outgoing edges in each DAG. Since weights of edges in each DAG (i.e., data to be transferred between tasks) are different, we generate files with different sizes to represent the weights on edges. Finally, the broker logs the execution cost of each IoT application and all of its constituent tasks in terms of selected evaluation metrics.
						
			\subsubsection{Baseline algorithms}
			We evaluate the performance of the $X$-DDRL with a greedy heuristic algorithm, and two DRL-based techniques from the literature that proposed DRL-based solutions for DAG-based IoT applications. In what follows, we briefly describe how these techniques are implemented, while their detailed specifications are provided in Section~\ref{relatedw}.   
			
			\begin{itemize}
%				\item \emph{Local}: In this method, all tasks of workflows are executed locally on their respective IoT devices, and hence, no parallel execution of tasks can be performed for workflows. The results of this method can be used as a reference point to analyze the gain of application placement techniques.
%				
%				\item \emph{Only Edge}: In this method, all tasks of workflows are offloaded to the fog/edge servers in the edge layer for the execution. If the VMs of all servers are full and there is no free VMs, the remaining tasks have to wait until free computing resources become available.     
%				
%				\item \emph{Only Cloud}: In this method, all tasks of workflows are executed on the cloud servers.      
%				
			\item \emph{PPO-RNN}: It is the extended and adapted version of the technique proposed in \cite{wang2020fast}\footnote{https://github.com/linkpark/metarl-offloading}. We extended this technique so that it can be used in multi-objective scenarios to minimize the weighted cost of execution. Besides, this technique is extended to be used in heterogeneous fog computing environments where several IoT devices, FSs, and CSs are available. This technique uses PPO as its DRL framework while the networks of the agent are wrapped by the RNN. Besides, we used the same hyperparameters as \cite{wang2020fast}. 
				
			\item \emph{PPO-No-RNN}: This technique is the same as PPO-RNN, while the networks are not wrapped by the RNNs.
				
			\item \emph{Double-DQN}: Many works in the literature uses standard Deep Q-Learning (DQN) based RL approach such as \cite{lu2020optimization, gazori2020saving, huang2019deep,xiong2020resource}. We implemented the optimized Double-DQN technique with an adaptive exploration for application placement in heterogeneous fog computing environments\footnote{https://docs.ray.io/en/master/}. The hyperparameters of this technique are set based on \cite{lu2020optimization}, which is a DQN-based application placement technique for DAG-based IoT applications.    
				
			\item \emph{Greedy}: In this technique, tasks are greedily assigned to the servers if their execution cost is less than the estimated local execution cost, similar to \cite{wang2020fast}.
					
			\end{itemize}
			
			\begin{figure*}[!ht]
			\begin{minipage}{\linewidth}
				\centering
				\begin{subfigure}{.325\textwidth}
					%\vspace{-2.5cm}
					%	\hspace{-0.5mm}
					\centering
					\includegraphics[width=\linewidth, height=4cm]{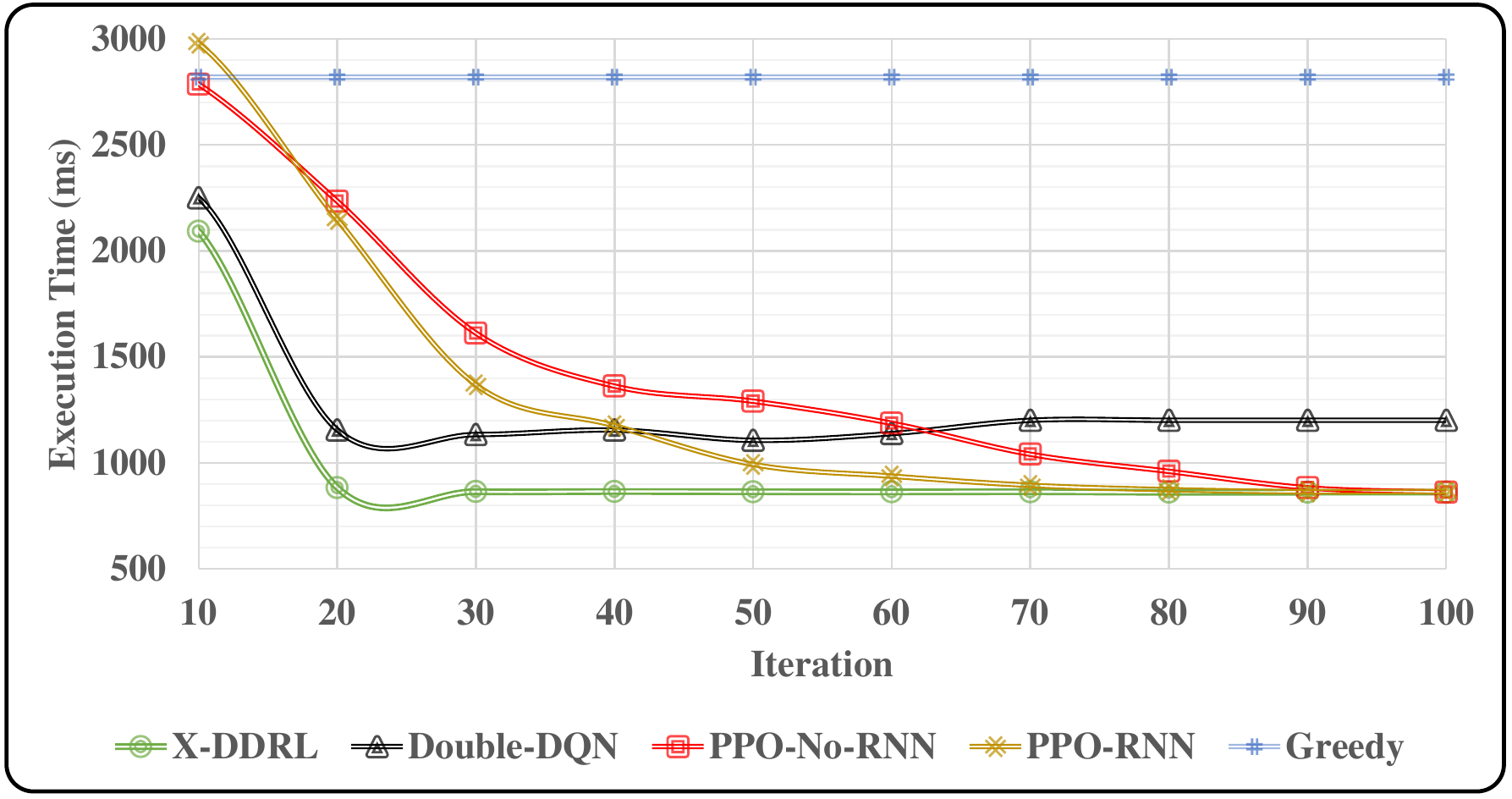}
					\captionsetup{justification=centering}
					\subcaption{Scenario 1: Execution time}
					\label{fig:convergenceAnalysis:sub1}
				\end{subfigure}%
				\hspace{0.25mm}
				\begin{subfigure}{.325\textwidth}
					%\vspace{0.2cm}
					\centering
					\includegraphics[width=\linewidth, height=4cm]{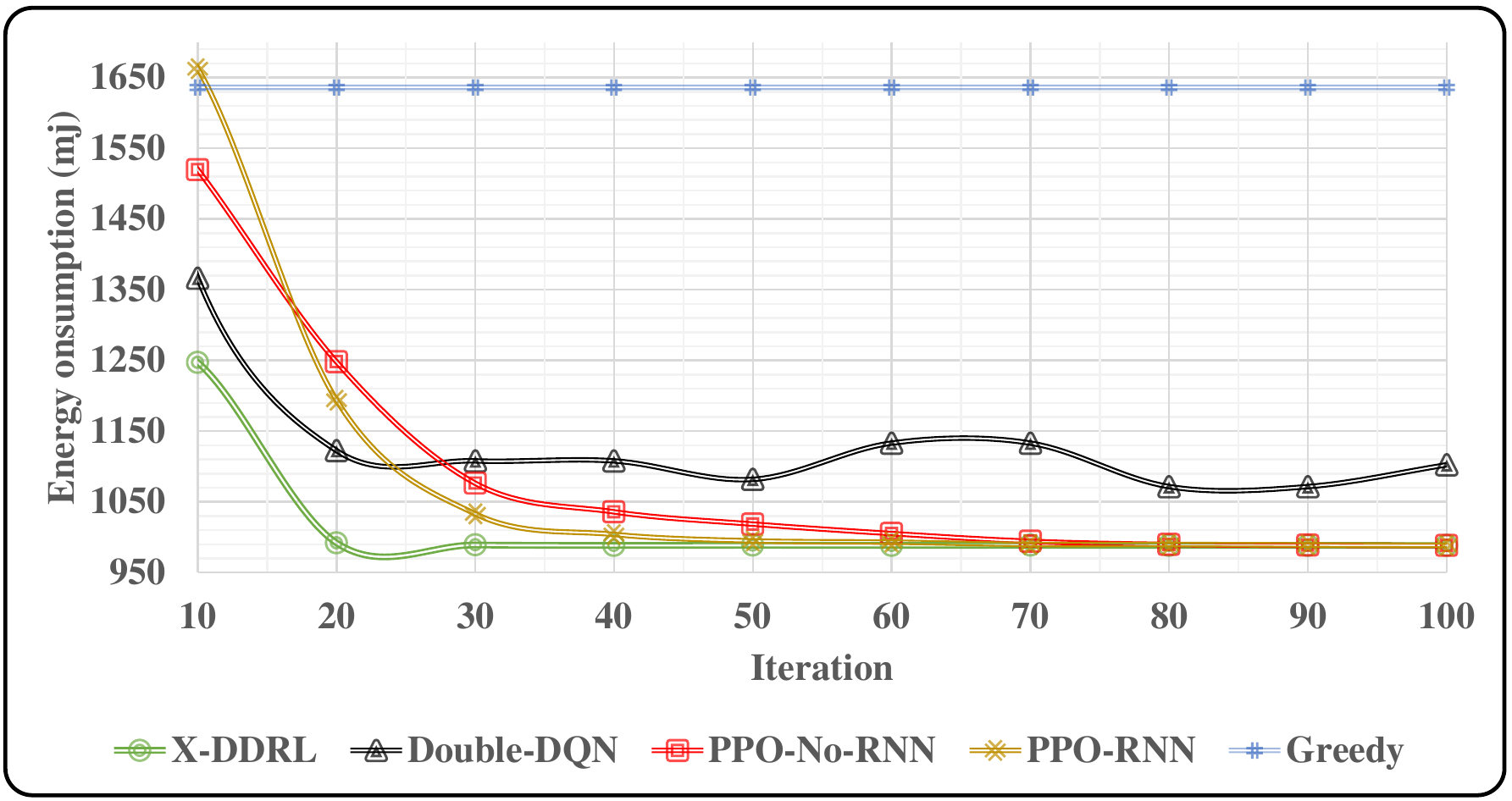}
					\captionsetup{justification=centering}
					\subcaption{Scenario 1: Energy consumption}
					\label{fig:convergenceAnalysis:sub2}
				\end{subfigure}%
				\hspace{0.25mm}
				\begin{subfigure}{.325\textwidth}
					%\vspace{0.2cm}
					\centering
					\includegraphics[width=\linewidth,height=4cm]{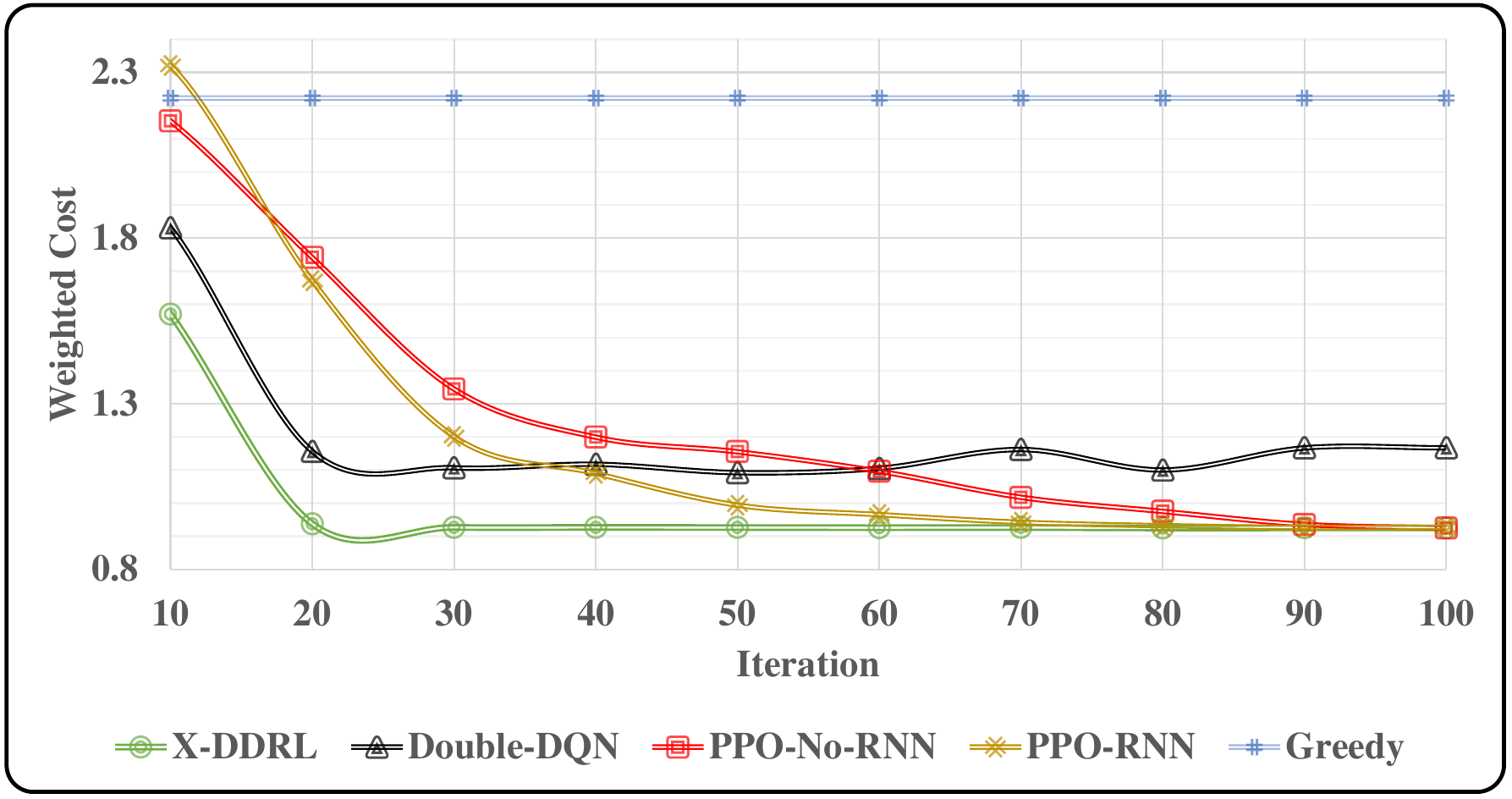}
					\captionsetup{justification=centering}
					\subcaption{Scenario 1: Weighted cost}
					\captionsetup{justification=centering}
					\label{fig:convergenceAnalysis:sub3}
				\end{subfigure}%
				\\
				\begin{subfigure}{.325\textwidth}
					%\vspace{-2.5cm}
					\centering
					\includegraphics[width=\linewidth, height=4cm]{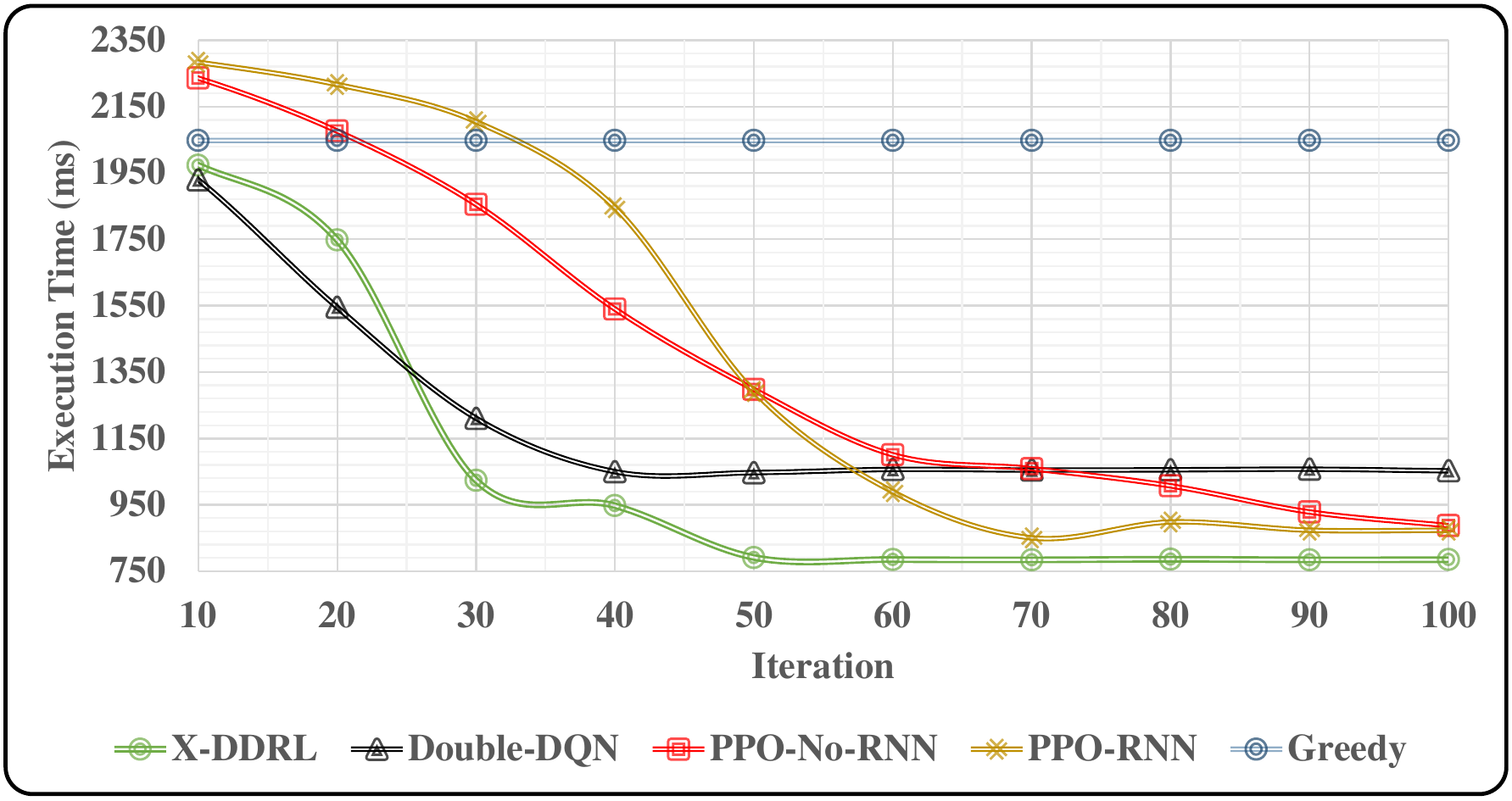}
					\captionsetup{justification=centering}
					\subcaption{Scenario 2: Execution time}
					\label{fig:convergenceAnalysis:sub4}
				\end{subfigure}
				\begin{subfigure}{.325\textwidth}
					%	\vspace{0.2cm}
					\centering
					\includegraphics[width=\linewidth,height=4cm]{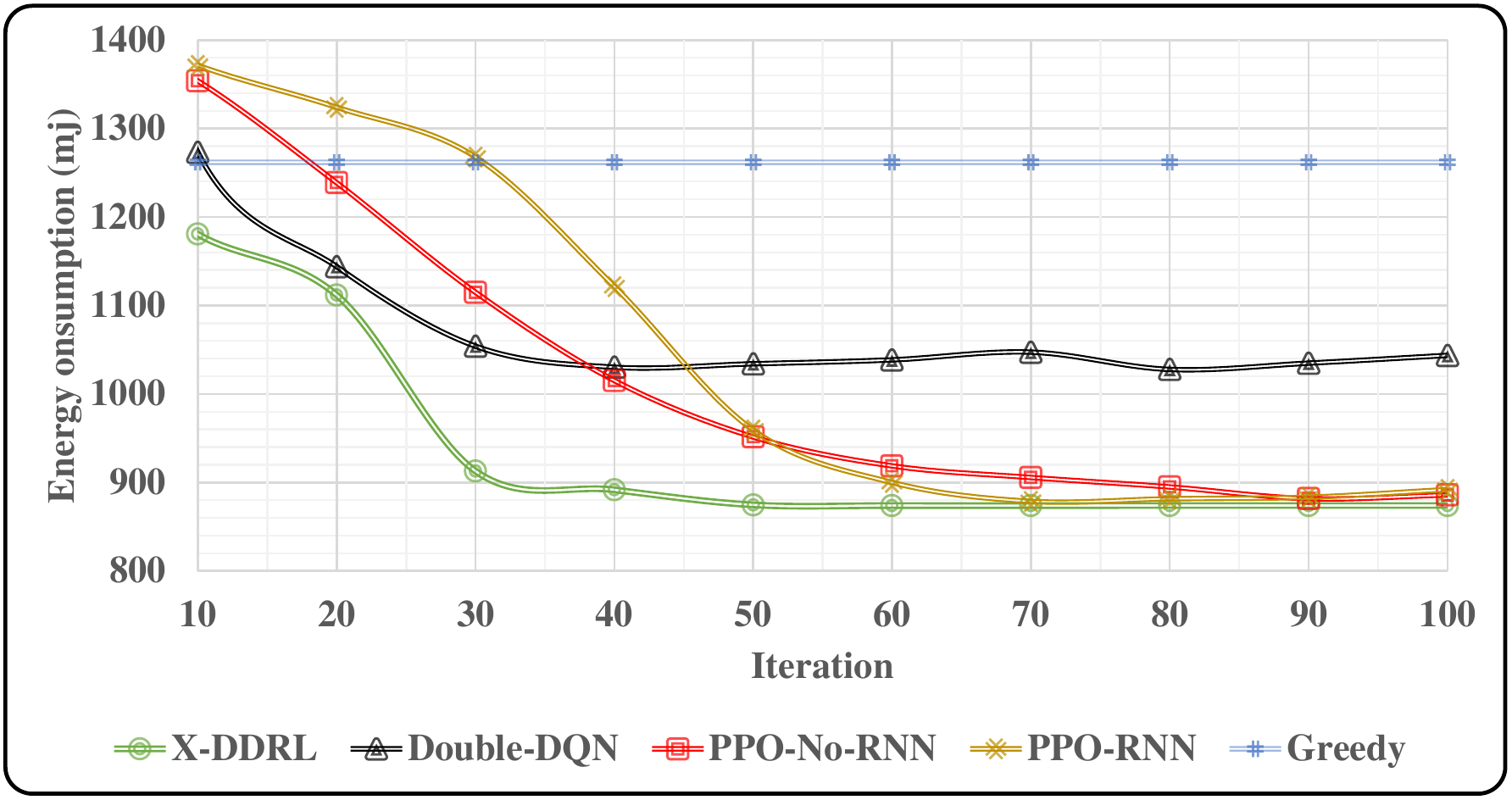}
					\captionsetup{justification=centering}
					\subcaption{Scenario 2: Energy consumption}
					\label{fig:convergenceAnalysis:sub5}
				\end{subfigure}
				\begin{subfigure}{.325\textwidth}
					%	\vspace{0.2cm}
					\centering
					\includegraphics[width=\linewidth, height=4cm]{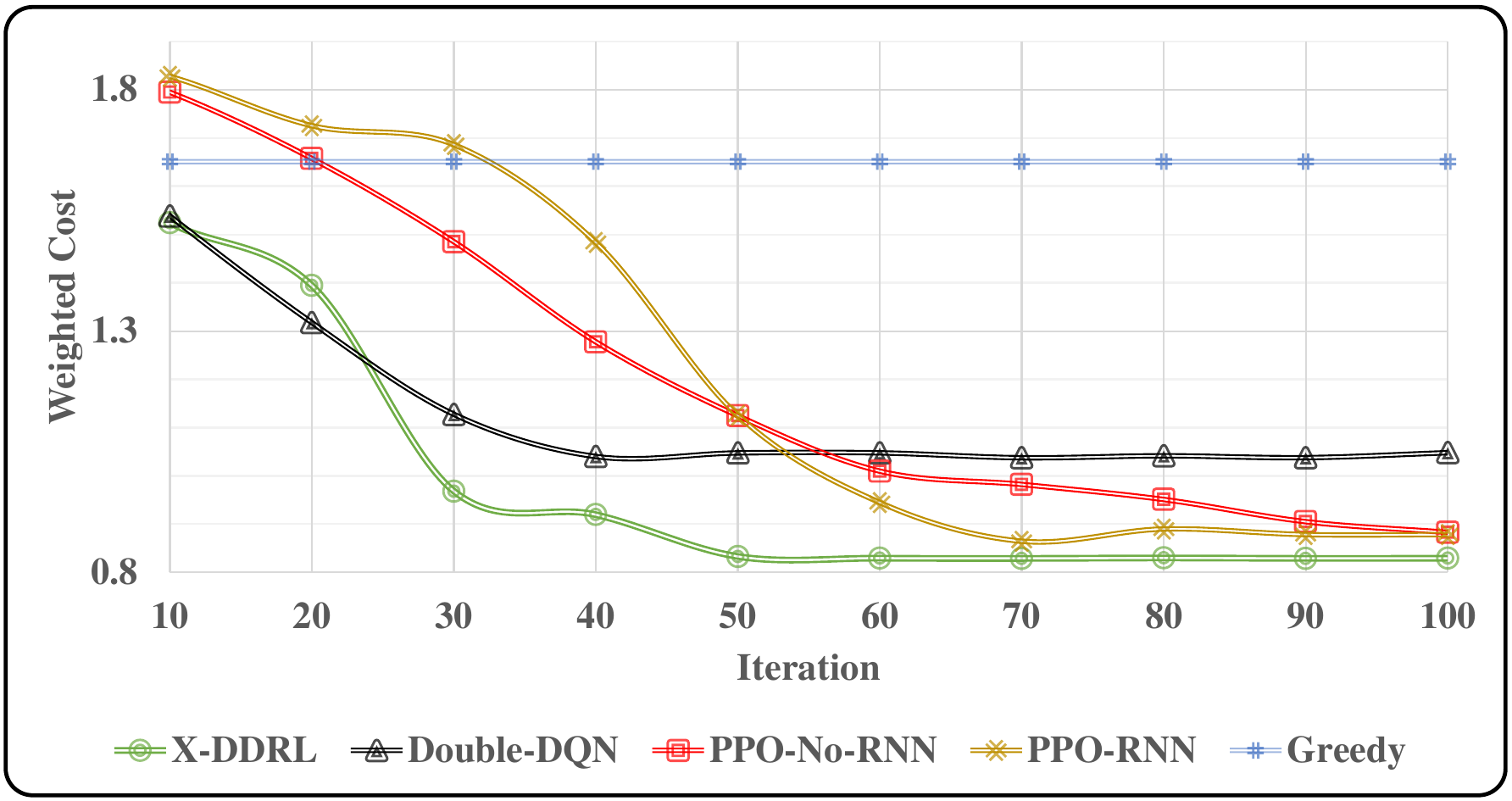}
					\captionsetup{justification=centering}
					\subcaption{Scenario 2: Weighted cost}
					\label{fig:convergenceAnalysis:sub6}
				\end{subfigure}

				\caption{Execution cost vs policy update analysis: In scenario 1, the training and evaluations are performed on datasets where $L=30$. In scenario 2, the training is performed on datasets where $L \in \{10,15,25,30\}$ and the evaluation is performed on datasets where $L=20$.}
				\label{fig:convergenceAnalysis}
			\end{minipage}
		\end{figure*}
		
		\subsection{$X$-DDRL Hyperparameters}
		In the implementation of $X$-DRRL, where the standard implementation of IMPALA is used\footnote{https://docs.ray.io/en/master/}, the DNN structure of all agents is similar, consisting of two fully connected layers followed by two LSTM layers as recurrent layers. Moreover, we performed a grid search to tune hyperparameters. According to tuning experiments, we set the learning rate $lr$ to $0.01$, the discount factor $\gamma$ to $0.99$. Besides, values of $\overline{\rho}$ and $\overline{c}$, controlling the performance of V-trace are set to $1$ \cite{espeholt2018impala} to obtain the best result. Table \ref{tab:hyperparameters} summarizes the setting of hyperparameters.
		
		\begin{table}[!t]
			\footnotesize
			\caption{The DNN and training hyperparameters}
			\centering
			\label{tab:hyperparameters}
			\renewcommand{\arraystretch}{1.2}
			\begin{tabular}{|c|c|c|c|}
				
				\hline
				\textbf{Parameter}     & \textbf{Value} & \textbf{Parameter}                               & \textbf{Value} \\ \hline
				Fully Connected layers & 2              & Learning Rate $lr$                               & 0.01           \\ \hline
				LSTM Layers            & 2              & Discount Factor $\gamma$                         & 0.99           \\ \hline
				Optimization Method    & Adam           & V-trace $\overline{\rho}$ & 1              \\ \hline
				Activation Function    & Tanh           & V-trace $\overline{c}$                    & 1              \\ \hline
			\end{tabular}
		\end{table}

		\subsection{Performance Study}
		In this section, four experiments are conducted to evaluate and compare the performance of $X$-DDRL with other techniques in terms of weighted execution cost, execution time of IoT applications, and energy consumption of IoT devices. 
	
		\subsubsection{Execution cost vs policy update analysis}
		\label{sec:convergenceAnalysis}
		In this experiment, we study the performance of application placement techniques in different iterations of the policy updates. We consider two scenarios for datasets of IoT applications to analyze how efficiently these techniques can extract features of different datasets of IoT applications and optimize their target policy. In the first scenario, we consider the number of tasks within IoT applications $L=30$. Hence, 25 datasets of IoT applications with the same task number and different $fat$ and $density$ are used, among which 20 datasets are used for the training and 5 datasets are used for the evaluation. In the second scenario, for the training $L \in \{10,15,25,30\}$ while for the evaluation $L=20$. Therefore, the training and evaluation are performed on datasets with a different number of tasks. Fig~\ref{fig:convergenceAnalysis} shows the obtained results of this study in terms of the average execution time of IoT applications, the energy consumption of IoT devices, and weighted cost for the above-mentioned two scenarios.
		\par
		As Fig.~\ref{fig:convergenceAnalysis} shows, the average execution cost of all techniques, except the greedy, decreases in different scenarios as the iteration number increases. However, the $X$-DDRL converges faster and to better placement solutions in comparison to other techniques. This is mainly because the V-trace function embedded in the $X$-DDRL uses n-step state-value approximation rather than 1-step state-value approximation \cite{espeholt2018impala}, improving convergence speed of $X$-DDRL to better solutions. Moreover, trajectories generated by distributed brokers are diverse, leading to a more efficient learning process. The execution cost of the greedy technique is fixed and does not change with different iteration numbers, but it can be used as a baseline technique to compare the performance of DRL-based techniques. The convergence speed of PPO-RNN and PPO-NO-RNN techniques is slower than the Double-DQN technique however, they finally converge to better placement solutions. In addition, the obtained results of the second scenario (Fig.~\ref{fig:convergenceAnalysis:sub4}, ~\ref{fig:convergenceAnalysis:sub5}, ~\ref{fig:convergenceAnalysis:sub6}) shows that all DRL-based techniques has lower convergence speed in comparison to the obtained results of first scenario (Fig.~\ref{fig:convergenceAnalysis:sub1}, ~\ref{fig:convergenceAnalysis:sub2}, ~\ref{fig:convergenceAnalysis:sub3}). However, still $X$-DDRL outperforms other techniques in terms of execution time, energy consumption, and weighted cost. This proves that the $X$-DDRL can more efficiently adapt itself with different DAG structures (i.e., task numbers, and dependency model), and hence it makes better application placement decisions in unforeseen scenarios.

		\subsubsection{System size analysis}
		In this experiment, the effect of different numbers of servers on application placement techniques is studied. The number of candidate servers has a direct effect on the complexity of application placement problems because the larger number of servers leads to a bigger search space. Hence, to analyze the performance of $X$-DDRL, the default number of servers in this experiment is multiplied by two and four; i.e, we have 24 and 48 servers respectively. Moreover, in this experiment, the training and evaluation datasets are specified as the same as the first scenario in Section~\ref{sec:convergenceAnalysis}; i.e., a total of 25 datasets where $L=30$ and different $fat$ and $density$ values. Due to the space limit and the fact that patterns for execution time, energy consumption, and weighted cost were roughly the same, only the obtained results from the weighted cost are provided in this experiment.    
		\par
\begin{figure}[!t]
			\centering 
			%\captionsetup{justification=centering,margin=2cm}
			%	\hspace{0cm}
			\includegraphics[height=5.5cm, width=\linewidth]{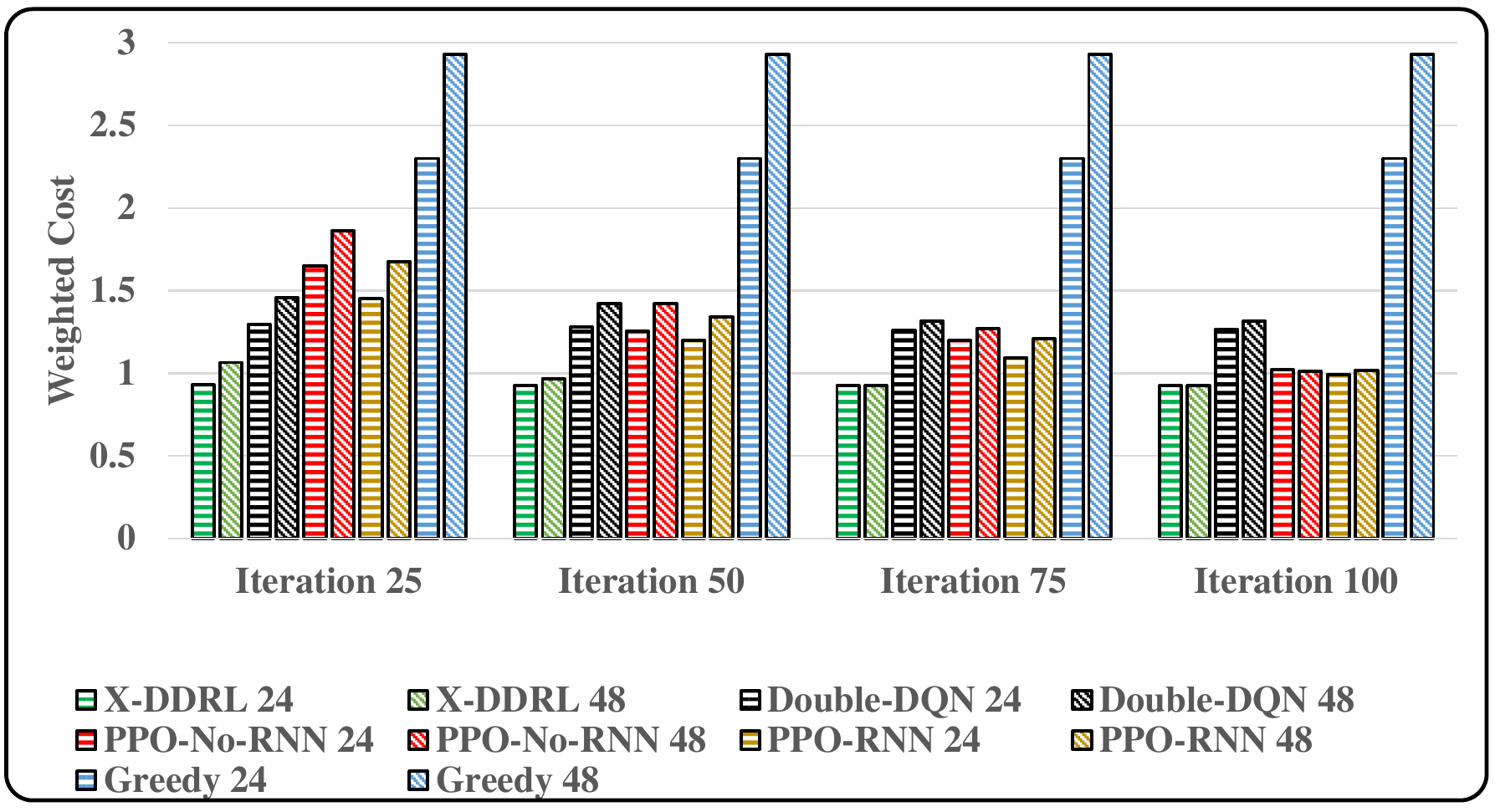}
			\caption{System size analysis}
			\label{fig:SystemSizeAnalysis}
			\vspace{-.4cm}
		\end{figure}

		Fig. \ref{fig:SystemSizeAnalysis} shows the weighted cost of different techniques, where brokers in the system have access to 24 and 48 candidate servers when making application placement decisions. It is crystal clear that the weighted cost of the greedy technique is steady for 24 and 48 servers as the number of iterations increases. All DRL-based techniques perform better than greedy technique either when the number of servers is 24 or 48. Also, it can be seen that the weighted execution costs of techniques are higher when the number of servers is 48 than weighted costs when the servers' number is 24. As the number of iterations increases, the DRL-based techniques can more accurately make placement decisions, leading to less weighted execution cost. However, the $X$-DDRL always outperforms other techniques and converges faster to better solutions. It shows that the $X$-DDRL has better scalability when the system size grows. This helps $X$-DDRL to make better application placement decisions in a fewer number of iterations. Among other DRL-based techniques, PPO-RNN performs better than PPO-No-RNN and Double-DQN and makes better placement decisions as the iteration numbers increases.
		
		\begin{figure}[!t]
			\centering 
			%\captionsetup{justification=centering,margin=2cm}
			%	\hspace{0cm}
			\includegraphics[height=5cm, width=\linewidth]{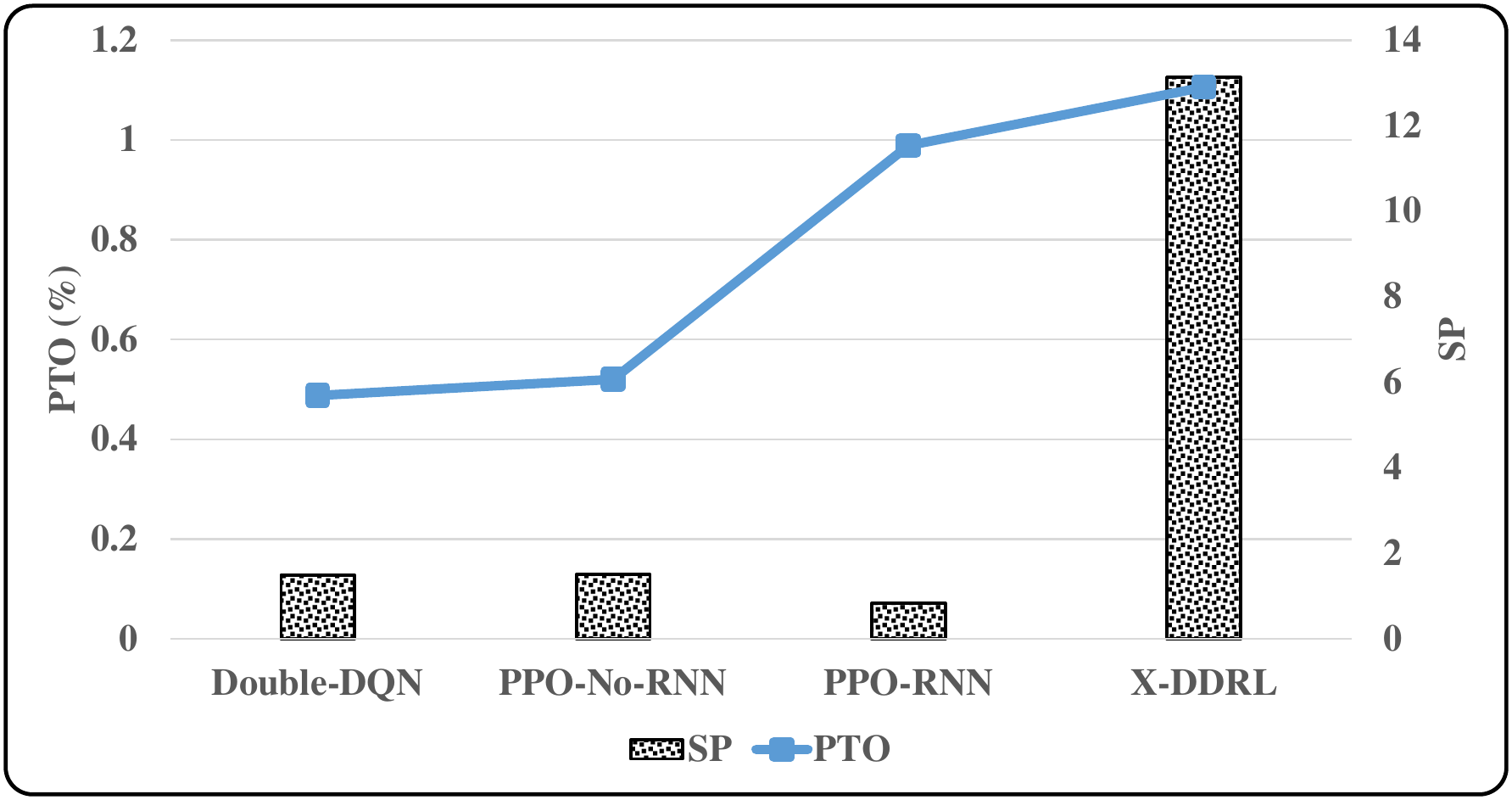}
			\caption{Placement time overhead and speedup analysis}
			\label{fig:SchedulingOverheadvsSpeedup}
		%	\vspace{-0.4cm}
		\end{figure}

			\begin{figure*}[!ht]
			%\begin{minipage}{\linewidth}
			\begin{subfigure}{.325\textwidth}
				\centering
				\includegraphics[width=6cm,height=4cm]{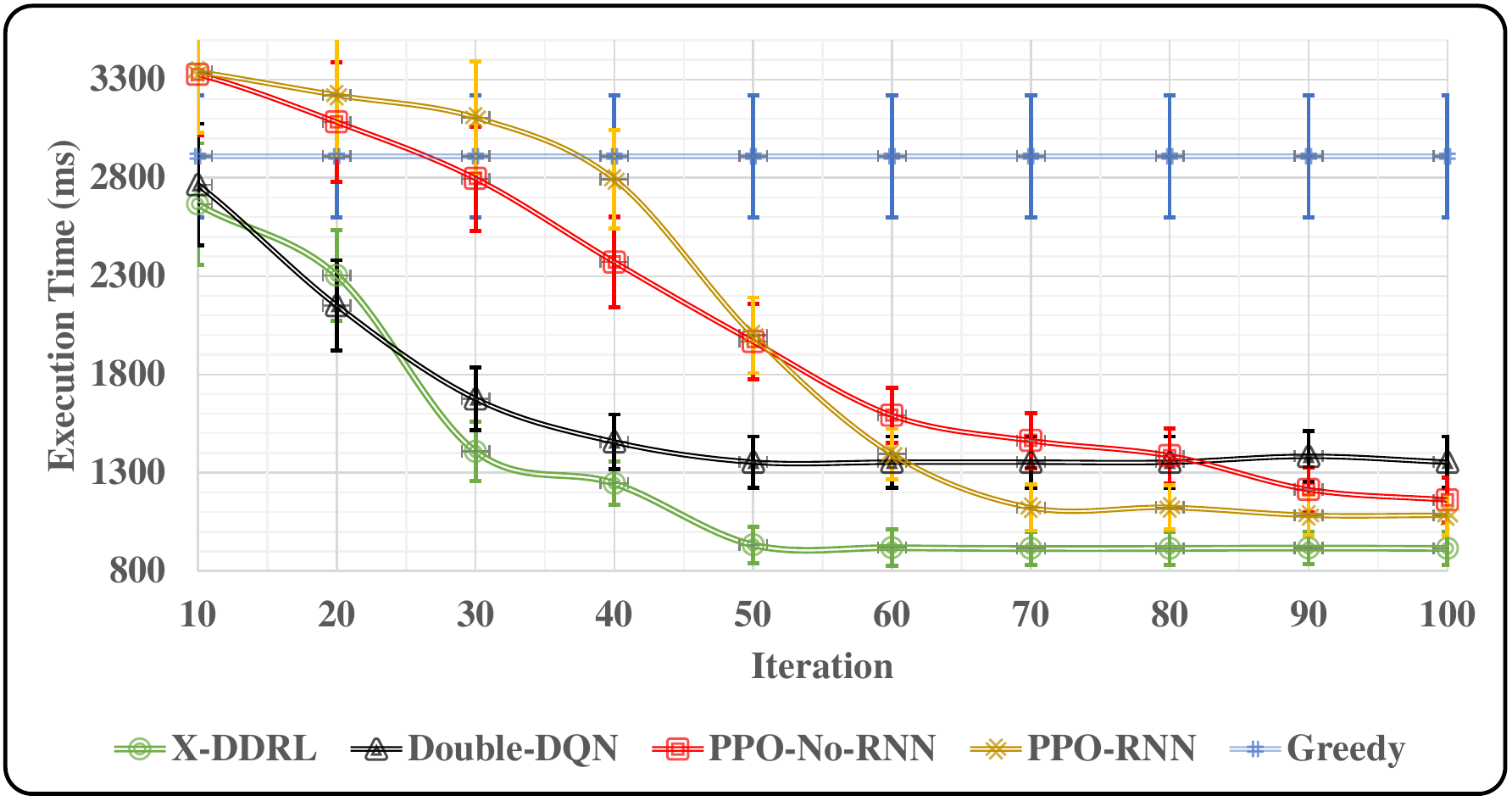}
				\captionsetup{justification=centering}
				\caption{Execution time}
				\label{fig:testbedResults:sub1}
			\end{subfigure}%
			\hspace{.05cm}
			\begin{subfigure}{.325\textwidth}
				\centering
				\includegraphics[width=6.05cm,height=4cm]{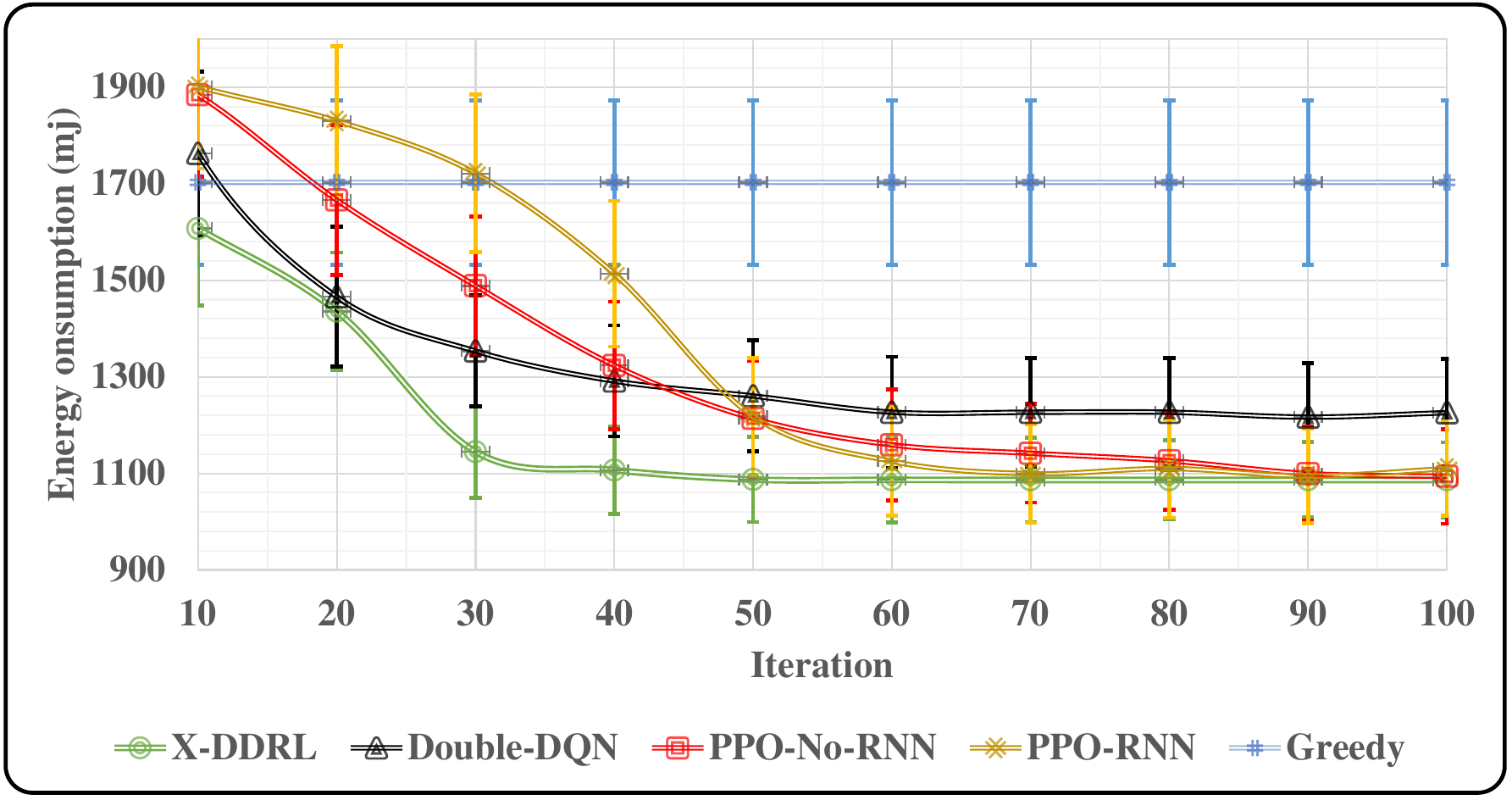}
				%\captionsetup{options}
				\captionsetup{justification=centering}
				\caption{Energy consumption}
				\label{fig:testbedResults:sub2}
			\end{subfigure}
			\hspace{.03cm}
			\begin{subfigure}{.325\textwidth}
				%	\vspace{0.2cm}
				\centering
				\includegraphics[width=6cm,height=4cm]{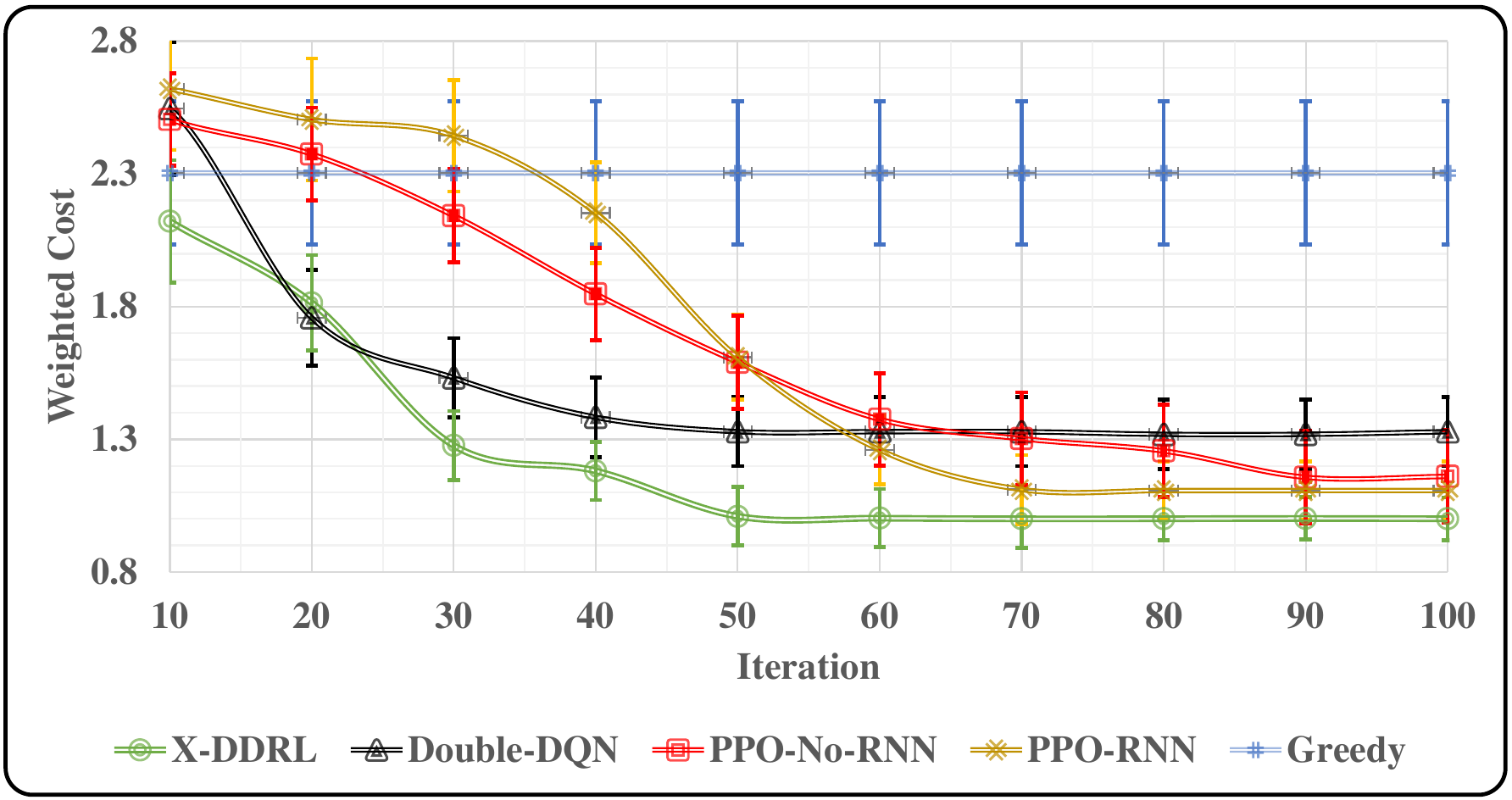}
				\captionsetup{justification=centering}
				\caption{Weighted cost}
				\label{fig:testbedResults:sub3}
			\end{subfigure}
			
			\caption{Evaluation on testbed}
			\label{fig:testbedResults}
		\end{figure*}

		\subsubsection{Speedup and placement time overhead analysis}
		In this section, we study the speedup and placement time overhead of different DRL-based techniques. We follow the same experimental setup as the first scenario in Section~\ref{sec:convergenceAnalysis}. We define the average Placement Time Overhead (PTO) as the average required amount of time for each technique to make an application placement decision divided by the average local execution time of IoT applications on IoT devices. To obtain the local execution time of IoT applications on IoT devices, we assume that tasks within an IoT application are executed sequentially, similar to \cite{xu2019computation}. Besides, we define the time taken by the $X$-DDRL technique with one broker to reach the value $1.1$ from the weighted execution cost as $Time_R$. The reason why $1.1$ is considered as the reference weighted execution cost is that this value is the minimum weighted execution cost that all DRL-based techniques can obtain. Moreover, the time taken by each technique to reach the reference weighted execution cost is defined as $Time_{T}$. Accordingly, similar to \cite{tuli2020dynamic}, the Speedup value of each technique ($SP$) is defined as $SP=\frac{Time_R}{Time_T}$. 
		\par
		%\vspace{-1mm}
		Fig~\ref{fig:SchedulingOverheadvsSpeedup} shows results of $PTO$ and $SP$ for all DRL-based techniques. The placement time overhead of techniques using RNN (i.e., $X$-DDRL and PPO-RNN) is usually higher than techniques that do not use RNN (i.e., Double-DQN, and PPO-No-RNN). The $PTO$ of the $X$-DDRL is higher than other DRL-based techniques by less than 1\% in the worst-case scenario, which is not significantly large. However, the obtained results of $SP$ show that $X$-DDRL performs 8 to 16 times faster than other techniques. Hence, considering the speedup performance and execution cost results of the $X$-DDRL, its placement time overhead is negligible, and $X$-DDRL can more efficiently perform application placement decisions compared to other techniques for heterogeneous fog computing environments.

		\subsubsection{Evaluation on Testbed}
To evaluate the performance of $X$-DDRL in real-world scenarios, we conducted experiments on the testbed whose configuration is discussed earlier in Section~\ref{sec:real-testbed}. In this experiment, for the training $L \in \{30,35,45,50\}$ while for the evaluation $L=40$.
\par
Fig~\ref{fig:testbedResults} shows the execution cost of different techniques in terms of execution time, energy consumption, and weighted cost by 95\% confidence interval. It can be observed that, similar to the simulation results, $X$-DDRL can outperform other techniques in terms of execution time, energy consumption, and weighted cost. Moreover, even after 100 iterations, where all techniques converged, there are no techniques that obtain better results in comparison to $X$-DDRL. It demonstrates that not only does $X$-DDRL converge faster, and its training time is significantly less than other techniques, but it also provides better results. As the results depict, the optimized Double-DQN technique converges faster than PPO-RNN and PPO-No-RNN, but it cannot obtain results as well as them. Overall, compared to converged results of other DRL-techniques, achieved results of $X$-DDRL show an average performance gain up to 30\%, 11\%, and 24\% in terms of execution time, energy consumption, and weighted cost, respectively.

\section{Conclusions and Future Work}
\label{conclusion}
This paper proposes a distributed DRL-based technique, called $X$-DDRL, to efficiently solve the application placement problem of DAG-based IoT applications in heterogeneous fog computing environments, where edge and cloud servers are collaboratively used. First, a weighted cost model for optimizing the execution time and energy consumption of IoT devices with DAG-based applications in heterogeneous fog computing environments is proposed. Besides, a pre-scheduling phase is used in the $X$-DDRL, by which dependent tasks of each IoT application are prioritized for execution based on the dependency model of the DAG and their estimated execution cost. Moreover, we proposed an application placement phase, working based on the IMPALA framework for the training of distributed brokers, to efficiently make application placement decisions in a timely manner. Distinguished from existing works, the $X$-DDRL can rapidly converge well-suited solutions in heterogeneous fog computing environments with a large number of servers and users. The effectiveness of $X$-DDRL is analyzed through extensive simulation and testbed experiments while comparing with the state-of-the-art techniques in the literature. The obtained results indicate that $X$-DDRL performs $8$ to $16$ times faster than other DRL-based techniques. Besides, compared to other DRL-based techniques, it achieves a performance gain up to 30\%, 11\%, and 24\% in terms of execution time, energy consumption, and weighted cost, respectively.  
\par  
As part of future work, we plan to extend our proposed weighted cost model to consider other aspects such as monetary cost, dynamic changes of transmission power, and total system cost. Moreover, we plan to apply mobility models in this scenario and adapt our proposed application placement technique accordingly.
			
		\section*{Acknowledgment}
		We thank Shashikant Ilager, Amanda Jayanetti, and Samodha Pallewatta for their comments on improving this paper.
			\vspace{-0.3cm}
			\bibliographystyle{IEEEtran}
			\bibliography{ref}
			\vspace*{-1.4cm}
			\begin{IEEEbiography}[{\includegraphics[width=1.05in,height=2in,keepaspectratio]{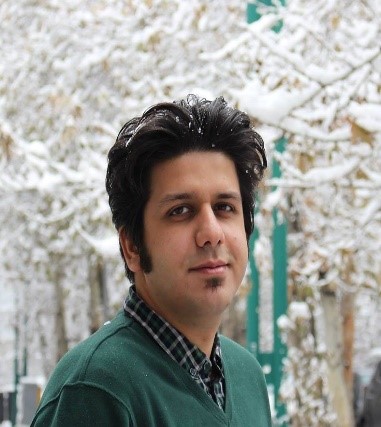}}]{Mohammad Goudarzi}
				 is working towards the Ph.D. degree at the Cloud Computing and Distributed Systems (CLOUDS) Laboratory, Department of Computing and Information Systems, the University of Melbourne, Australia. He was awarded the Melbourne International Research Scholarship (MIRS) supporting his studies. His research interests include Internet of Things (IoT), Fog Computing, Wireless Networks, and Distributed Systems.
			\end{IEEEbiography}
		
		\vspace{-1.1cm} 
		\begin{IEEEbiography}[{\includegraphics[width=1.05in,keepaspectratio]{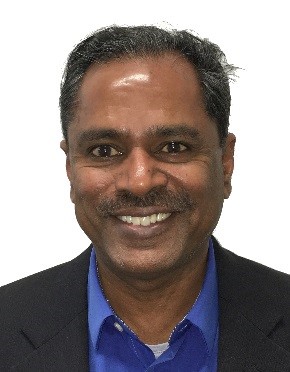}}]{Marimuthu Palaniswami} is a Fellow of IEEE and past distinguished lecturer of
			the IEEE Computational Intelligence Society. He received his Ph.D. from the University of Newcastle, Australia before joining the University of Melbourne, where he is a Professor of Electrical Engineering. Previously, he was a Co-Director of Centre of Expertise on Networked Decision \& Sensor Systems. He has published more than 500 refereed journal and conference papers, including 3 books, 10 edited volumes.
			
		\end{IEEEbiography} 
		\vspace{-1.2cm} 
		\begin{IEEEbiography}[{\includegraphics[width=1.05in,keepaspectratio]{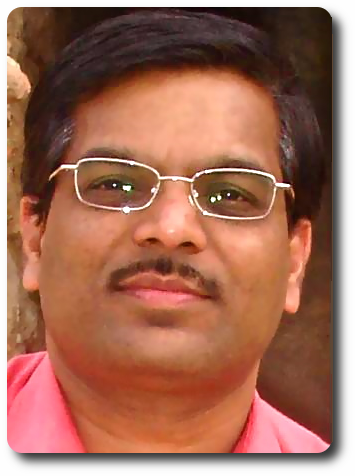}}]{Rajkumar Buyya} is a Redmond Barry Distinguished Professor and 
			Director of the Cloud Computing and Distributed Systems (CLOUDS) 
			Laboratory at the University of Melbourne, Australia.
			He has authored over 625 publications and seven text books including
			"Mastering Cloud Computing" published by McGraw Hill, China Machine 
			Press, and Morgan Kaufmann for Indian, Chinese and international markets 
			respectively.  He is one of the highly cited authors in computer science
			and software engineering worldwide (h-index=149, g-index=322, 116,000+ 
			citations).
		\end{IEEEbiography}

		\end{document}